\documentclass[preprint2]{aastex}

\shorttitle{1H~0323+342: {\it Swift} Monitoring and {\it Suzaku} Observation}

\shortauthors{Su Yao et al.}

\usepackage{amssymb}
\usepackage{amsmath}
\usepackage{natbib}
\usepackage{multirow}

\usepackage[usenames]{color}

\newcommand{\LAT}{{\it Fermi}/LAT }

\begin{document}


\title{The $\gamma$-ray Detected Narrow-Line Seyfert 1 Galaxy 1H~0323+342:
{\it Swift} Monitoring and {\it Suzaku} Spectroscopy}


\author{	Su Yao\altaffilmark{1}, 
		Weimin Yuan\altaffilmark{1}, 
		S. Komossa\altaffilmark{2,1}, 
		Dirk Grupe\altaffilmark{3,4},
		L. Fuhrmann\altaffilmark{2}, 
		\& 
		Bifang Liu\altaffilmark{1}}


\altaffiltext{1}{National Astronomical Observatories, Chinese Academy of Sciences, 20A Datun Road, Chaoyang District, Beijing, China; yaosu@nao.cas.cn, wmy@nao.cas.cn}
\altaffiltext{2}{Max-Planck Institut f{\"u}r Radioastronomie, Auf dem H{\"u}gel 69, 
53121 Bonn, Germany; skomossa@mpifr-bonn.mpg.de, lfuhrmann@mpifr-bonn.mpg.de}
\altaffiltext{3}{Space Science Center, Morehead State University, Morehead, KY, 40351, 606-783-9597}
\altaffiltext{4}{Swift Mission Operation Center, 2582 Gateway Dr. State College, PA, 16801, USA; dgrupe007@gmail.com}


\begin{abstract}
As a radio-loud narrow-line Seyfert 1 galaxy (NLS1) detected by \LAT in GeV $\gamma$-rays, 
1H~0323+342 is a remarkable Active Galactic Nucleus (AGN) showing properties characteristic of 
both NLS1s and blazars. 
Here we present results of simultaneous  X-ray and UV/optical monitoring observations 
on 1H~0323+342 taken with the UV/Optical Telescope (UVOT) and X-ray Telescope (XRT) onboard the {\it Swift} satellite 
over six years from 2006. 
Overall, the object showed statistically correlated variations in both the UV and X-ray bands on timescales of years as well as on timescales of days. 
A deep {\it Suzaku} observation reveals X-ray variability on timescales 
as short as a few tens of thousand seconds, and an X-ray spectrum typical of  Seyfert galaxies. 
The broad-band spectral energy distribution, for which the data of UV and X-ray observations taken on 2009 July 26--27 were used, can be well modeled with a simple one-zone leptonic jet  model 
plus  accretion disk/corona emission. 
The latter is predominantly responsible for the
 UV/optical and  X-ray (0.3--10\,keV) emission and their observed variations. 
The  correlated UV--X-ray variability on the timescale of days 
is consistent with reprocessing of the  X-ray radiation by the accretion disk. 
The shortest timescale  and large normalized excess variance of the X-ray variability detected with {\it Suzaku}
suggest a relatively small black hole mass of the order of $10^7M_{\odot}$, 
consistent with the estimation based on the broad H$\beta$ line in the optical. 

\end{abstract}


\keywords{galaxies: active --- galaxies: individual (1H~0323+342) --- galaxies: perticular --- galaxies: Seyfert}

\section{INTRODUCTION}

As a subclass of Active Galactic Nuclei (AGNs), 
narrow-line Seyfert 1 galaxies (NLS1s) are thought to have black holes with lower masses and higher Eddington ratios compared to normal Seyfert 1s (\citealt{1992ApJS...80..109B, 2004ApJ...606L..41G, 2004AJ....127.3168B, 2012AJ....143...83X}; see \citealt{2008RMxAC..32...86K} for a review). 
While 15\%-20\% of normal AGNs are radio-loud \citep{1995PASP..107..803U}, 
the fraction seems to be much smaller in NLS1s. 
Using the Sloan Digital Sky Survey Data Release 3 (SDSS-DR3), \citet{2006ApJS..166..128Z} 
 built a large sample of $\sim2000$ NLS1s, in which only $\sim$7\% 
were detected by the FIRST survey \citep[the Faint Images of the Radio Sky at 20 cm survey;][]{1995ApJ...450..559B}. 
In a systematic search for radio-loud NLS1 galaxies, \citet{2006AJ....132..531K} reported a fraction of 7$\%$ radio-loud NLS1s, while only 2.5\% are very radio loud (radio loudness $R_{5\rm GHz}>100$).
\citet{2008ApJ...685..801Y}  analyzed 23 genuinely radio-loud NLS1s with the loudness\footnote{The radio loudness $R_{1.4\rm GHz}\equiv f(1.4{\rm GHz})/f(4400{\rm \AA})$ in \citet{2008ApJ...685..801Y} is defined as the ratio between the flux densities at 1.4~GHz and 4400\AA\, in the rest frame, instead of commonly used $R_{5\rm GHz}\equiv f(5{\rm GHz})/f(4400{\rm \AA})$ \citep[][]{1989AJ.....98.1195K}. 
}
$R_{1.4\rm GHz}>100$ and found that some of the objects show observational properties
 characteristic of blazars with relativistic jets. 
This finding was confirmed and highlighted by the later detections of 
 GeV $\gamma$-ray emission with the {\it Fermi}/LAT
\citep{2009ApJ...699..976A, 2009ApJ...707L.142A, 2012MNRAS.426..317D},
as well as the rapid variability in the infrared and optical bands 
in three of the extreme objects, which are also detected in $\gamma$-rays \citep{2010ApJ...715L.113L, 2012ApJ...759L..31J}. 
Besides all the above properties, the high brightness temperatures \citep[see, e.g.,][]{2003ApJ...584..147Z, 2006PASJ...58..829D,2007ApJ...658L..13Z,2008ApJ...685..801Y,2011A&A...528L..11G,2012ApJ...760...41D} and the superluminal motion measured from the radio images \citep[see, e.g.,][]{2012MNRAS.426..317D, 2013MNRAS.436..191D} also point to the presence of the relativistic jets. 
Given the peculiarity of NLS1s among AGNs
(i.e.\ relatively small black hole masses, high accretion rates, possibly at early phases of the AGN evolution), 
the interesting question arises as to whether the jetted NLS1s are simply downsized extensions of blazars, 
or whether they form a new population. 

Among the several $\gamma$-ray detected NLS1s, 1H~0323+342 (RA=$3^{\rm h}24^{\rm m}41^{\rm s}$.161, DEC=$34^{\circ}10'45''.86$, \citealt{2002ApJS..141...13B}; also known as 2MASX J03244119+3410459, \citealt{2006AJ....131.1163S}) is the nearest one with a redshift of $z=0.0629$ \citep{2007ApJ...658L..13Z}, allowing detailed observational studies. 
It was first discovered  as an X-ray source in the {\it HEAO-1} X-ray survey \citep{1984ApJS...56..507W}
and  was spectroscopically identified with 
 a Seyfert 1 galaxy in the optical with strong Fe{\scshape~ii} lines and weak forbidden lines  \citep{1993AJ....105.2079R}. 
Its hybrid nature sharing properties of both NLS1s (e.g., a small broad line width of 1600\,km\,s$^{-1}$, strong Fe{\scshape~ii} emission lines)
and blazars was first reported by \citet{2007ApJ...658L..13Z}, 
making it a prototype of this newly recognized class of AGNs.
The radio source has a flat radio spectrum with an index $\alpha_r=+0.1$ ($S_\nu\propto\nu^{\alpha_r}$) \citep{1994A&AS..106..303N, 2012ApJ...760...41D}, and a high radio loudness $R_{5\rm GHz}=246$ \citep{2012ApJ...760...41D} or $R_{1.4\rm GHz}=318$ \citep{2011nlsg.confE..24F}. 
The VLA 1.4~GHz map revealed a core with a two-sided structure of  $\sim$15kpc
\citep{2008A&A...490..583A}. 
The brightness temperatures $T_{\rm B}$ estimated from both the 
radio images and flux variations are larger than $5\times10^{10}$\,K, exceeding the upper limit  expected from energy equipartition \citep{2014ApJ...781...75W}. 
All these observations suggest the presence of a relativistic jet in this object. 
The optical image taken with the 
{\it Hubble Space Telescope (HST)} revealed 
a one-armed spiral structure \citep{2007ApJ...658L..13Z} of the host galaxy of 1H~0323+342, 
whereas a ring-like morphology was suggested based on ground-based observations
 \citep{2008A&A...490..583A}. 
The asymmetric structure of the host galaxy revealed by the analysis of imaging data leads to the suggestion that 1H~0323+342 may be associated with a merger \citep{2008A&A...490..583A, 2014ApJ...795...58L}. 
Based on the broad H$\beta$ line width and luminosity, a black hole mass of  $1.8\times10^{7}M_{\odot}$ was suggested by \citet{2007ApJ...658L..13Z}, with an uncertainty of 0.3--0.4 dex as commonly inferred \citep{2006ApJ...641..689V, 2009ApJ...707.1334W, 2012ApJ...755..167D}.

At high energies, its 0.3--10\,keV spectrum with a low signal-to-noise ratio (S/N) obtained by {\it Swift}/XRT can be  fitted by a power law with a 
photon index of $\Gamma\approx2$ \citep{2007ApJ...658L..13Z}, 
whereas the spectrum becomes flatter in the range 20--100\,keV, 
$\Gamma=1.55$, as obtained by {\it INTEGRAL} \citep{2011MNRAS.417.2426P}. 
Its X-ray emission is highly variable \citep[e.g.][]{2007ApJ...658L..13Z, 2009ApJ...707L.142A, 2014arXiv1405.0715P}. 
The  detection of 1H~0323+342 with the {\it Fermi}/LAT \citep{2009ApJ...707L.142A} is remarkable,
confirming the presence of  a relativistic jet. 
The non-thermal jet emission can account for the two bumps of 
its broad-band spectral energy distribution (SED), 
one in the radio--infrared band and the other one in the GeV $\gamma$-ray band. 
In addition, an accretion disk component is also present \citep{2009ApJ...707L.142A}. 
Interestingly, variations of the hard X-ray emission above 20\,keV were found by \citet{2009AdSpR..43..889F}: 
 the source had a low flux and a soft spectrum as observed with 
{\it INTEGRAL}/IBIS in 2004, 
and a  high flux and a hard spectrum as observed with {\it Swift}/BAT in 2006--2008. 
This was interpreted as being dominated by emission from the disk/corona during the former observation, and by a possible contribution from the jet flaring in the hard X-ray band above 20\,keV during the latter observation. 

1H~0323+342 has been monitored by the {\it Swift} satellite \citep{2004ApJ...611.1005G}
in both  the X-ray and  the UV/optical bands simultaneously over a period of more than six years from 2006 to 2013. 
 It  was also observed  with the {\it Suzaku} X-ray satellite \citep{2007PASJ...59S...1M} with a relatively long exposure on 2009 July 26--27. 
These observations provide valuable datasets to study the X-ray and UV/optical 
radiation of this object and their relationship. 
In this work we analyze the long-term X-ray and UV/optical data of 1H~0323+342 
to study both the temporal and spectral properties and their variations in these wavebands. 
Of particular interest, we find statistical correlations of the emission between the X-ray and UV
bands on both long timescales of years and  relatively short timescales of days. 
In Section~\ref{reduction}, the observations and data reduction are described. 
The temporal and spectral analysis are presented in Section~\ref{timing} and 
Section~\ref{spectrum}, respectively, followed by the broad-band SED. 
A summary of the results, and their implications are presented and discussed in Section~\ref{discussion}. 
Errors on parameters of spectral modeling are quoted at 
the 90\% confidence level unless mentioned otherwise. 
Throughout this paper a cosmology is assumed with $H_0=70$\,km\,s$^{-1}$\,Mpc$^{-1}$, $\Omega_\Lambda=0.73$ and $\Omega_{\rm M}=0.27$.

\section{OBSERVATIONS AND DATA REDUCTION}
\label{reduction}

\subsection{The {\it Swift} Data\label{swift}}

\begin{figure}
\epsscale{1.20}
\plotone{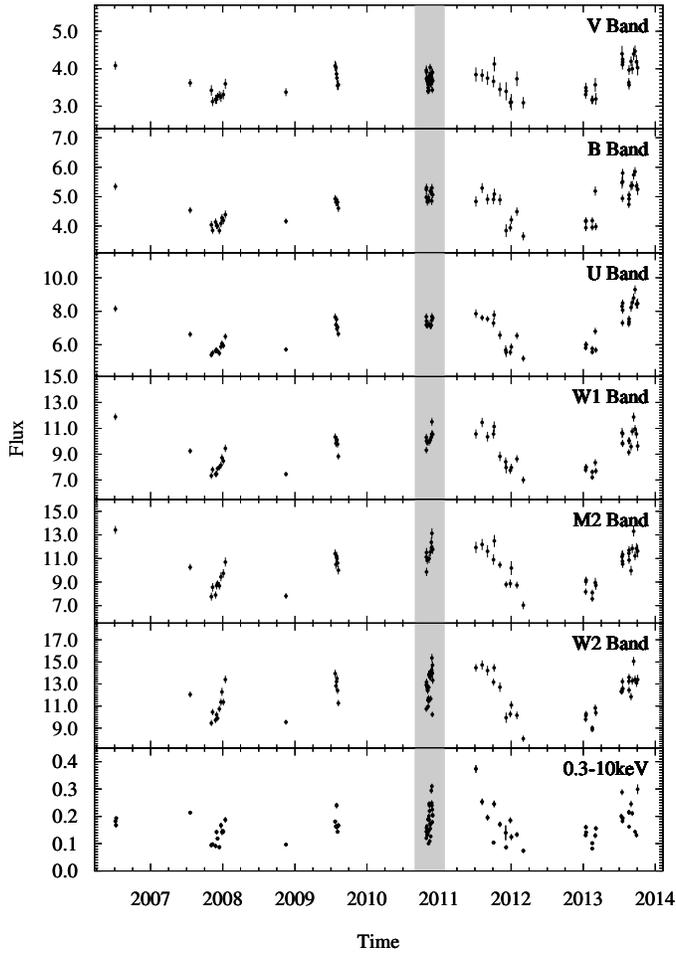}
\caption{{\it Upper six panels}: Light curves in the $v$, $b$, $u$, $w1$, $m2$, $w2$ band, respectively, of {\it Swift}/UVOT from 2006 to 2013. 
The fluxes are in units of $10^{-16}$\,erg\,s$^{-1}$\,cm$^{-2}$\,\AA$^{-1}$. 
{\it Bottom panel}: X-ray count rates in the band 0.3--10\,keV observed by the {\it Swift}/XRT. 
The observations in 2010 October--November (in the grey area) are used to investigate the correlations of the UV and X-ray variability on short timescales. \label{swiftlc}}
\end{figure}

The {\it Swift} satellite has monitored 
1H~0323+342 
at 84 occasions 
from 2006 July to 2013 October, with  data from the X-ray Telescope \citep[XRT, 0.3--10\,keV,][]{2005SSRv..120..165B} and/or the UV/Optical Telescope \citep[UVOT, 1700\AA-6000\AA,][]{2005SSRv..120...95R}. 
Some of the observations were performed at relatively high cadence 
with timescales of $\sim$1 day (as shaded in Figure~\ref{swiftlc}). 
We have retrieved data from the public {\it Swift} archives, 
including observations published in \citet{2009AdSpR..43..889F} and \citet{2014arXiv1405.0715P}. 
All data are reduced using the {\it Swift} {\tt FTOOLS} in {\tt HEASOFT V6.12}. 

The XRT observations in the photon-counting mode \citep[PC,][]{2004SPIE.5165..217H} are used and the data
are reduced using  the task {\tt xrtpipeline V0.12.6}. 
X-ray events are selected with grades 0--12 and extracted using {\tt xselect V2.4}. 
To check whether the data suffered from the piled-up effects, 
0.3--10\,keV X-ray spectra are first extracted from a circle of 47$''$ radius (enclosing 90\% of the 
point spread function (PSF) at 1.5\,keV) centered on the source on the observed X-ray images. 
Some spectra have count rates higher than 0.5 counts\,s$^{-1}$, indicating that
 they may be affected by pile-up\footnote{http://www.swift.ac.uk/analysis/xrt/pileup.php}. 
To eliminate the effects of pile-up, the central region of the source images on the
 detector are excluded.
The size of the exclusion region is determined by fitting a King function \citep{2005SPIE.5898..360M}
 to the source image of the data with the highest count rates, and an inner exclusion radius
 of 10$''$ \citep[see also][]{2010ApJ...716...30A} is found. 
Thus a source extraction region  avoiding  possible pile-up pixels 
is defined to be an annulus with an inner radius of 10$''$ and an outer radius of 70$''$.
For constructing light curves, 
source counts are extracted from such an annulus centered at the source position 
for all the data\footnote{
We have examined the possible effect of the uncertainty of the source position
(which is better than 3$''$ \citep[$\sim$1~pixel,][]{2005SSRv..120..165B})
on the extracted source counts by offsetting the center of the source extraction region
by one pixel in various directions from the nominal source center. 
It is found that the resulting differences in the extracted source counts are at a level of
 only a few percent at most, which are negligible compared with the flux variations of the source
 (see below).
 };
while for extracting  source spectra, it is only applied to those data suffering from pile-up.
The background region is chosen to be an annulus of the same area centered on the source with 
an inner radius of 75$''$ and an outer radius of 102$''$. 
The spectra are re-binned using {\tt grppha V3.0.1} to have at least 20 counts per bin. 
The ancillary response function files are created by {\tt xrtmkarf} and corrected for vignetting, hot columns and bad pixels using the exposure maps. 
We use the relevant respond matrix given in the output of {\tt xrtmkarf}. 

During most of the {\it Swift} observations, the object was also observed with 
UVOT in all the six passbands in the optical ($v$, $b$, $u$ bands) and near UV ($w1$, $m2$, $w2$ bands). 
Images of the same band during each observation are stacked after aspect correction. 
The source region is chosen to be the standard aperture of a circle with a 5$''$ radius, and the background is estimated from a circle with a 15$''$ radius near the source region without contamination from other sources \citep[see also][]{2010ApJ...716...30A}. 
The photon counts are first converted into fluxes and then magnitudes in each band using the conversion factors with the task \texttt{uvotsource} in the {\it Swift} data analysis package \citep{2008MNRAS.383..627P}, which was obtained by assuming a GRB-like power-law spectrum with dust extinction. 
However, the $b-v$ colors derived are out of the range valid for this conversion factor, indicating a different spectral shape from the one assumed as default.    
We thus calculate a new conversion factor (as well as the effective wavelength and Galactic extinction) for each of the passbands applicable to our target using its actual spectral shape. 
This is done by folding the spectrum with the effective area of each of the filters
of UVOT following Raiteri et al. (2010). The observed spectrum is determined by fitting the average UV/optical photometric data of 1H~0323+342 with a log-parabolic model. 
The Galactic extinction curve from Fitzpartick (1999) is employed with $E(B-V)=0.2095$ for this object taken from Schlegel et al. (1998). 
The calibration results are shown in Table~\ref{swift_uvot}. 
Then the de-reddened fluxes of 1H~0323+342 in the UVOT bands are calculated using the new conversion factors derived above.   

\subsection{The {\it Suzaku} Data\label{suzaku}}

The X-ray Imaging Spectrometer \citep[XIS, 0.2--12\,keV,][]{2007PASJ...59S..23K} 
onboard the {\it Suzaku} satellite observed 1H~0323+342 on 2009 July 26--27 
(ObsID: 704034010, PI: Hayashida, M.) for about two days 
with a total net exposure of $\sim$84~ks. 
Available data are provided by the front-illuminated (FI) CCDs XIS0, XIS3, and the back-illuminated (BI) CCD XIS1 (XIS2 was lost in 2006). 
We use the {\it Suzaku} {\tt FTOOLS} version 19 and associated {\tt CALDB} released on 2013 March 5. 
The data are first re-calibrated and screened using {\tt aepipeline}. 
Generally an extraction circle of 260$''$ radius encircling 99\% of a point source flux is recommended. 
In order to keep enough region for the background extraction as well as to compensate the lost photons in the wings of the PSF, 
the source light curves and spectra are extracted from a source-centered circle of 210$''$ radius, 
and the background light curves and spectra are extracted from a source-free circle of the same radius, avoiding the chip corners containing the calibration sources. 
To increase the S/N, we combine light curves of all the CCDs,
and also light curves of  observations in both the 3$\times$3 and 5$\times$5 modes,
into a single light curve with bins of 256 s. 
For each CCD exposure, the  data are examined 
and found to be free of  pile-up effects \citep{suzakupileup}
by checking the PSF of the X-ray images and the count rates  at the image peak.
The Redistribution Matrix Files (RMF) and Ancillary Response Files (ARF) are generated by using the tasks {\tt xisrmfgen} and {\tt xissimarfgen}, respectively. 
Then the spectra from XIS0 and XIS3 are combined by using {\tt addascaspec} to increase the S/N, 
whereas the spectrum from XIS1 is considered separately since the BI CCD has a distinctly different response from the FI CCDs. 
The spectra are grouped to have a minimum counts of 200 per bin. 
Channels in the 1.7--2.3\,keV range are ignored because of the possible uncertain calibration of an instrumental Si K edge at $\sim$1.84\,keV, a K$\alpha$ fluorescence line at $\sim$1.74\,keV and an Au M edge at $\sim$2.2\,keV\footnote{http://heasarc.gsfc.nasa.gov/docs/Suzaku/analysis/abc/}. 
The spectral analysis is performed by using 
 {\tt XSPEC V12.7.1} \citep{1996ASPC..101...17A} and the $\chi^2$ minimization technique is applied. 


We also analyze the data from the PIN silicon diodes and GSO phoswich 
counters of the Hard X-ray Detector \citep[HXD, 10--600\,keV,][]{2007PASJ...59S..35T}.
The source was not significantly detected above 20\,keV, however, because of the high background. 
The spectra from PIN and GSO are extracted automatically by using the tasks {\tt hxdpinxbpi} and {\tt hxdgsoxbpi} after recalibration and screening. 
Then we re-bin the spectrum and exclude those with negative net counts. 
The upper limit on the hard X-ray fluxes is estimated in the following way. 
The observed  number of counts $C_{\rm t}$ in each bin is expected to follow the poisson statistics. 
So the expected number of counts is distributed as $Q(0.9, C_{\rm t})$ at the (one-tail) probability of 90\%. 
The  upper limit on the source counts 
at the 90\% confidence level can be estimated as $Q(0.9, C_{\rm t})-C_{\rm b}$, 
where $C_{\rm b}$ is the background counts \citep{2014ApJ...782...55Y}. 
The counts are converted to fluxes using {\tt Xspec} by assuming a power-law spectrum with $\Gamma=1.9$, similar to that at lower energy (see Section~\ref{x_spec}).

\section{VARIABILITY ANALYSIS}
\label{timing}

\subsection{Long-Term Variability}

\begin{figure}
\epsscale{0.90}
\plotone{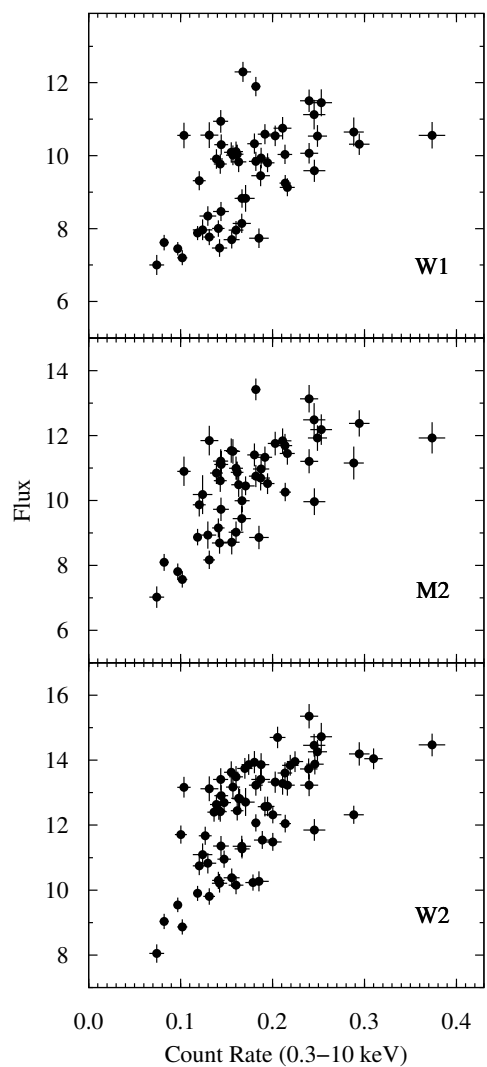}
\caption{The X-ray count rates versus UV fluxes for all the {\it Swift} observations analyzed in this paper. 
The UV fluxes are in units of $10^{-16}$\,erg\,s$^{-1}$\,cm$^{-2}$\,\AA$^{-1}$. \label{xrt_uvot_long}}
\end{figure}

The multi-wavelength variability of 1H~0323+342 is investigated 
by using the X-ray count rates and UV/optical fluxes derived from the {\it Swift} and {\it Suzaku} observations
between 2006 and 2013. 
The long-term light curves in various bands, of which each data point represents one observation,  are shown in Figure~\ref{swiftlc}. 
We test for the presence of significant variability at the UV/optical by using the $\chi^2$ test 
against the null hypothesis of no variation. 
The source is found to be significantly variable in essentially all bands with $p$-values of  $P<10^{-8}$. 
A trend is apparent 
that the UV and X-ray variability follow  each other on timescales as long as years, and it becomes less significant in the optical bands. 
To verify this correlated variations, 
the UV fluxes, which are obtained from the count rates as described in Section~\ref{swift}, 
are plotted against the X-ray count rates for the whole dataset in Figure~\ref{xrt_uvot_long}. 
Albeit with large scatter, the trend of correlated variations between the X-ray and the UV bands 
is evident. 
This is confirmed by the non-parametric Spearman Rank correlation test \citep{1992nrca.book.....P}, 
which gives extremely small values of the null probabilities ($p$-values are $P=2.8\times10^{-6}$, $P=1.3\times10^{-7}$ and $P<10^{-8}$ for the $w1$, $m2$ and $w2$ band respectively, see Table\,\ref{short_variability})
for the observed correlations arising by chance. 

\subsection{Short-Term Variability on Timescales of Days \label{variation}}

In this work we focus on the short-term variations on timescales of 
 days and less in 1H~0323+342. 
For this we make use of  data from the intensive monitoring campaign at high cadence performed  
 by {\it Swift}/XRT during 2010 October--November (shaded region in Figure~\ref{swiftlc}).
Moreover,  the {\it Suzaku} X-ray data taken with a long exposure  on 2009 July 26--27 are also used.  

For the UV/optical light curves during 2010 October--November, 
we first test the significance of variations using the $\chi^2$ test. 
The source is considered to be variable if the chance probability $P<0.1\%$.  
We find that the UV/optical variability of 1H~0323+342 is wavelength dependent. 
While no variability is detected in the $b$, $u$, and ultraviolet $w1$ band, the object is significantly variable in the $v$ band (with $P=1.2\times10^{-4}$) and in the two ultraviolet bands $m2$ and $w2$ (with $P=3.8\times10^{-4}$ and $P<10^{-8}$ for the $m2$ and $w2$ bands respectively). 
Two factors may account for this difference. 
First, it has been known that in AGNs the UV/optical variability amplitude increases with decreasing wavelength \citep[e.g.][]{1985ApJ...296..423C, 2004ApJ...601..692V, 2013AJ....145...90A}. 
Second, there is an increasing amount of the host galaxy starlight 
with increasing wavelength towards the optical, which dilutes the variations of the central AGN. 
This is particularly true for 1H~0323+342 for its prominent host galaxy
 as revealed by the optical {\it HST} image taken with the F702W filter (6919\AA), which 
extends out to at least $8''$  due to its proximity
and contributes about half of the total optical light of the whole galaxy \citep{2007ApJ...658L..13Z}. 
Such a size is larger than the extraction radius of $5''$ used in the photometry analysis above 
(Section~\ref{swift}). 
We check the {\it HST} snapshot images of 1H~0323+342. Unfortunately, we find that the images are saturated at the central point source (AGN) due to overexposure of the imaging observations, which hampers any reliable imaging analysis of the AGN-host decomposition. 
However, we expect that the contamination from the host galaxy light in the UV bands is rather low. 
This is because there is evidence that the central region of the host galaxy is dominated by a bulge \citep[][]{2014ApJ...795...58L}, and consistently, our new Palomar 5-m spectroscopic observations in the near-IR on this object reveal an old stellar population  (Zhou et al. in preparation). 
We thus conclude that the UV fluxes are essentially dominated by the AGN flux and use them in the variability and SED analysis below. 
%
A close-up of the UV and X-ray light curves in the time interval 2010 October--November is shown in Figure~\ref{w2_x}. 
For the UV, only the $w2$ band data are displayed, 
because the sampling with the other two filters is rather sparse. 
Significant variations in  the UV and particularly the X-ray bands are evident. 
Some interesting features can be seen. 
One may notice that a rapid drop and recovery of $\sim$0.4 mag within a single day at the epoch of day 30 is observed in the $w2$ light curve. 
Similarly rapid changes, of lower amplitudes, occur at other time intervals. 
We have carefully checked the images of those datasets for the aspect correction and other photometric issues, and confirm that the variations are real. 

In order to make direct comparison of the variability between the UV and X-ray bands,
we calculate the fractional variability amplitudes $F_{\rm var}$ for these bands \citep{2003MNRAS.345.1271V}, 
\begin{equation}
	\label{frac_var}
	F_{\rm var}=\frac{ \sqrt{\sigma^2-\langle\varepsilon^2\rangle} }
				{\langle f\rangle},
\end{equation}
where $\sigma^2$ is the variance of the light curve, 
$\langle\varepsilon^2\rangle$ the mean of the squared measurement errors, and
 $\langle f\rangle$ the mean flux.
The 1-$\sigma$ uncertainty of $F_{\rm var}$ due to measurement errors is estimated following \citet{2003MNRAS.345.1271V}. 
The results are given in Table~\ref{short_variability}, showing 
 a much larger fractional variability amplitude in the X-ray than in the UV band. 
 Another often used estimator of the variability amplitude is the normalized excess variance \citep{1997ApJ...476...70N, 2003MNRAS.345.1271V}, which is the square of $F_{\rm var}$ and conveys exactly the same information as $F_{\rm var}$.
We will talk about this estimation in Section~\ref{excessvariance}.

The X-ray variability on even shorter timescales is investigated using the {\it Suzaku} observations.
Figure~\ref{suzakulc}  shows the  {\it Suzaku}/XIS light curves in the total 0.2--12\,keV band (upper panel),
 and in the soft (0.2--2\,keV, middle panel) and hard band (2--12\,keV, lower panel). 
The light curves are modulated by the short orbital period for {\it Suzaku}'s low-altitude orbit (gaps between exposures).
It can be seen that during the {\it Suzaku} observation
the source varied dramatically  by a factor of about two
on a timescale as short as $\sim$20~ks. 
The variations seem to be independent of the energy bands during the {\it Suzaku} observation since the same variability is seen
in both the soft and hard band.

\subsection{Correlated Short-Term UV and X-ray Variations}
\label{correlated_uv_x}

\begin{figure}
\epsscale{1.20}
\plotone{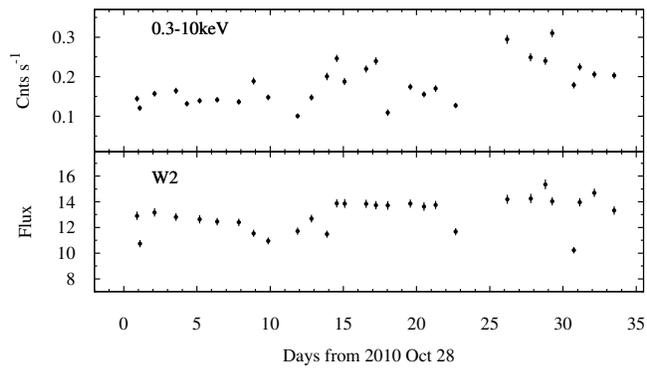}
\caption{Densely sampled X-ray and $w2$ light curves in 2010 October--November. 
The $w2$ flux is in units of $10^{-16}$\,erg\,s$^{-1}$\,cm$^{-2}$\,\AA$^{-1}$. 
A dip in both the X-ray and $w2$ band of 1H~0323+342 is present around the 10th day. 
\label{w2_x}}
\end{figure}

As indicated in Figure~\ref{w2_x},
the two light curves ($w2$ and X-ray) appear to show somewhat coordinated variation patterns.
A significant drop followed by a rise up within a few days occurred both in the $w2$ and 
X-ray bands around day 10 and this behavior appears to be simultaneous. 
However, the degree of correlated variations is much lower in the second half of the light curve. 
From day 15 to day 22, the X-rays  varied strongly by nearly a factor of two whereas the $w2$ light curve remained  stable. 
After day 25, both X-rays  and $w2$ varied, but the correlation between them seems not to be so obvious. 
There seems to be another drop in $w2$ from day 23 but its presence is not clear because of a large gap between observations. 

Since the data samplings in the two bands are nearly paired up during 2010 October--November, we can 
 examine any possible statistical correlations by using a correlation test.
Figure~\ref{xrt_uvot_short} shows the X-ray count rates versus the $w2$ fluxes
for each pair of the observations. 
The results of the Spearman correlation test are given in Table~\ref{short_variability}.
As can be seen, there is a statistically significant correlation between the two bands, 
with a null probability  $1\times10^{-5}$. 
Using the UV fluxes in the other two bands ($w1$, $m2$) also yields statistically significant  correlations between the UV and X-rays (Table~\ref{short_variability}),
although the significance is relatively lower due to the more sparse data points.

Given the  overall statistical correlations of the X-ray and $w2$ light curves, 
we search for any possible time lag between them. 
The cross-correlation function (CCF) method is used, which is commonly employed  to search for time lags 
between two sets of time series data in AGN studies  \citep{1992ApJ...386..473H, 2000ApJ...533..631K}. 
To handle the unevenly sampled light curves, two complementary methods are adopted here, 
 the interpolated cross-correlation function \citep[ICCF;][]{1986ApJ...305..175G, 1987ApJS...65....1G} with interpolation bins of 1 day and the  $z$-transformed discrete correlation function \citep[ZDCF;][]{1997ASSL..218..163A}. 
As indicated in Figure~\ref{ccf}, both methods give generally consistent results 
and reveal a broad signal peaking at  $\tau\sim0$~day
 (positive time lag means the X-rays are leading the UV).  
We fit the ZDCF in the $-6<\tau<9$ range
with a Gaussian profile and find it peaking at $\tau\approx0.3$~day.
A maximum likelihood method introduced by \citet{2013arXiv1302.1508A}
gives a peak at $\tau=0.6^{+2.7}_{-1.0}$~day with 1-$\sigma$ uncertainty. 
These results may indicate a correlation with a possible time lag around zero days between the two bands with X-ray tentatively leading. 
However, the uncertainty is large and the lag is statistically consistent with zero, given the broad peak of the CCF. 

To test the significance of the cross-correlation function, we simulate X-ray light curves that are uncorrelated with the observed one and then calculate the ICCFs between simulated X-ray light curves and the observed UV light curve. 
The detailed procedure of the simulations is described in Appendix~\ref{append}. 
The 68\%, 95\% and 99\% extremes of the ICCFs distribution out of the two uncorrelated light curves are indicated by dashed, dash-dotted and dotted lines in Figure~\ref{iccf_test}. 
As shown in the figure, the peak around zero lag in the real data reaches a significance level just slightly below 95\%, and thus can only be considered to be marginal. 
The lower significance level found for the correlation than that from the direct correlation analysis using the Spearman's Rank correlation test (Table~\ref{short_variability}) is mostly due to the random fluctuations introduced from the interpolation of the two light curves by assuming that the light curves vary smoothly \citep[e.g.][]{1994PASP..106..879W, 1997ASSL..218..163A}. 
We note that there appears to be a second peak around $\tau\sim-10$~days. 
However, given the even lower significance and the shortened segment of data involved in the correlation ($\sim25$~days), we do not consider this peak to be a real correlation, rather a coincidence of the second UV dip ($\sim$23~day, see Figure~\ref{w2_x}) with the first X-ray dip. 
In other parts of the whole long-term light curves, the number of paired data points with good sampling cadence is too small to carry out meaningful cross-correlation analysis.

\section{SPECTRAL ANALYSIS}
\label{spectrum}


\subsection{X-ray Spectral Analysis}
\label{x_spec}


The X-ray spectra observed with {\it Suzaku}/XIS are analyzed for their much higher S/N compared to 
the {\it Swift}/XRT spectra. The combined  FI  spectrum in the band 0.5--10\,keV is used.
We first consider the total, time-averaged spectrum, as shown in Figure\,\ref{averaged_spec}.
All the best-fit parameters are listed in Table~\ref{best_fit_averaged}. 
An absorbed single power-law model with the absorption fixed at the Galactic value \citep[$N^{\rm Gal}_{\rm H}=1.27\times10^{21}$\,cm$^{-2}$; ][]{2005A&A...440..775K} gives an unacceptable fit ($\chi^2$/d.o.f.=923/362), with a systematic excess of flux below 1\,keV in the residuals (Figure~\ref{averaged_spec} b). 
A black body ({\tt zbbody}, with a redshift fixed at $z=0.0629$) is then added to account for this soft excess, which improves the fit significantly
($\chi^2$/d.o.f.=354/360), 
giving a black body temperature $kT=0.15\pm0.01$\,keV and a photon index $\Gamma=1.87\pm0.02$. 
The temperature is within the previously observed range $\sim$ 0.1--0.2\,keV of the soft X-ray excess in the 
X-ray spectra of AGNs \citep{2004MNRAS.349L...7G, 2004A&A...422...85P, 2006MNRAS.365.1067C, 2011ApJ...727...31A}. 
The photon index is somewhat flat for  NLS1s but not unseen, because there is a large scatter in the distribution of spectral indices of NLS1s \citep[e.g.][]{1999ApJS..125..317L, 2010ApJS..187...64G, 2011ApJ...727...31A}. 
We also consider a double power-law model, which gives a good fit. 
Alternatively, to account for the soft X-ray emission, we fit a partial covering  model with ionized absorption
\citep[{\tt zxipcf} in {\tt Xspec},][]{2008MNRAS.385L.108R}, 
which results in a somewhat worse $\chi^2$ and a marginally acceptable fit 
with a lowly ionized absorber of $N_{\rm H}=6\times10^{22}$\,cm$^{-2}$ 
and a moderate covering fraction of $f=58\%$. 
We also fit a Comptonization model \citep[{\tt comptt} in {\tt Xspec},][]{1994ApJ...434..570T} and fix the seed photon energy at the innermost temperature of the standard accretion disk based on the black hole mass and accretion rate of this object (see below). 
The model results in an improved fit over the black body model, and the fitted electron temperature $kT_{\rm plasma}=0.28$\,keV and optical depth $\tau=13.4$ are within the previously observed range in similar objects \citep[e.g.][]{2004MNRAS.349L...7G, 2011ApJ...727...31A}.

In addition to the soft X-ray excess, there is a systematic excess in the 
residuals above 6\,keV (Figure~\ref{averaged_spec}c,d). 
We have also used a primary power-law continuum plus a reflection component. 
The ionized disk reflection model from \citet[{\tt reflionx} in Xspec]{2005MNRAS.358..211R} is employed. 
In order to account for the relativistic blurring caused by strong gravitation near the black hole, we use {\tt relconv} \citep{2010MNRAS.409.1534D} as a relativistic convolution kernel to convolve with the disk reflection component. 
The free parameters are the index of the primary power-law component $\Gamma$, the black hole spin $a$, the abundance of iron relative to the solar value $A_{\rm Fe}$ and the ionization parameter $\xi_{\rm R}$. 
A similar approach was adopted by \citet{2013MNRAS.428.2901W} to investigate the Suzaku observations of a sample of  AGNs including 1H 0323+342. 
They found that the inclination obtained was unphysically high for an unobscured active galaxy ($i=82^\circ\pm3^\circ$), so the inclination angle was fixed at 45$^{\circ}$ in the fitting. 
Here we fixed the inclination at $10^\circ$, considering the small inclination angle of the jet inferred from its blazar properties and detected $\gamma$-ray emission. 
The inner radius of the accretion disk is fixed at the innermost stable circular orbit and the outer radius fixed at $400r_{\rm g}$, where $r_{\rm g}$ is the gravitational radius. 
The radial emissivity profile of the disk is assumed to be a power law $\epsilon\propto r^{-q}$ with $q=3$. 
The model gives a very good fit ($\chi^2$/d.o.f.=351/355). 
The spin parameter, though poorly constrained, is found to be $<0.13$ at the 90\% confidence level. 
This result is different from that of \citet{2013MNRAS.428.2901W} possibly due to the very different inclination angle adopted here, but consistent with the finding of generally low black hole spins for NLS1s \citep{2015MNRAS.447..517L}. 
However, a reliable estimate on the black hole spin from fitting the reflection component can only be achieved by extending the observed X-ray spectrum to higher energies up to 30--40\,keV. 

Fitting the spectrum from the single BI CCD (XIS1) gives essentially consistent  results 
with those for the FI CCD spectra. 
Since adding the single BI spectrum results in only limited  improvement in the S/N,
we do not incorporate it in the analysis, considering possible systematics in the cross-calibration.

\begin{figure}
\epsscale{1.20}
\plotone{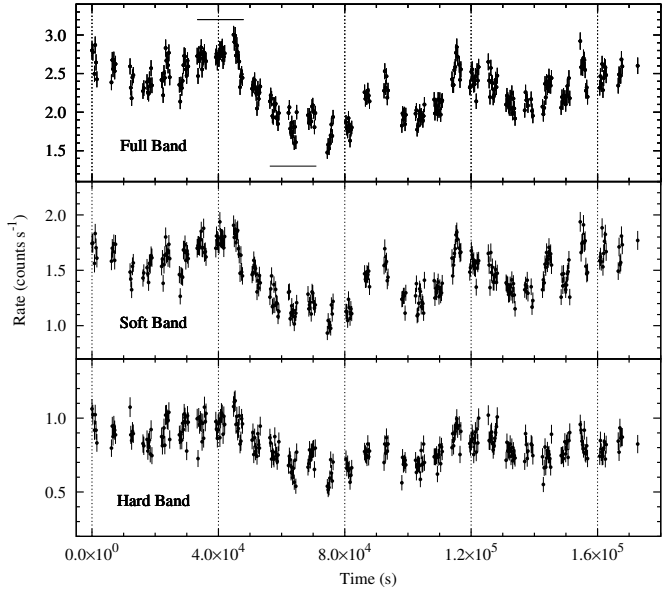}
\caption{The light curves of 1H~0323+342 in different energy bands observed by the {\it Suzaku}/XIS, from top to bottom: the full XIS band light curve (0.2--12\,keV), the soft band light curve (0.2--2\,keV) and the hard band light curve (2--10\,keV). The bars in the top panel cover the time ranges for which the high- and low-state spectra are extracted. The four 40~ks long segments which are used to calculate the normalized excess variance are separated by the vertical dashed lines (see Section~\ref{excessvariance}). \label{suzakulc}}
\end{figure}

\begin{figure}
\epsscale{1.20}
\plotone{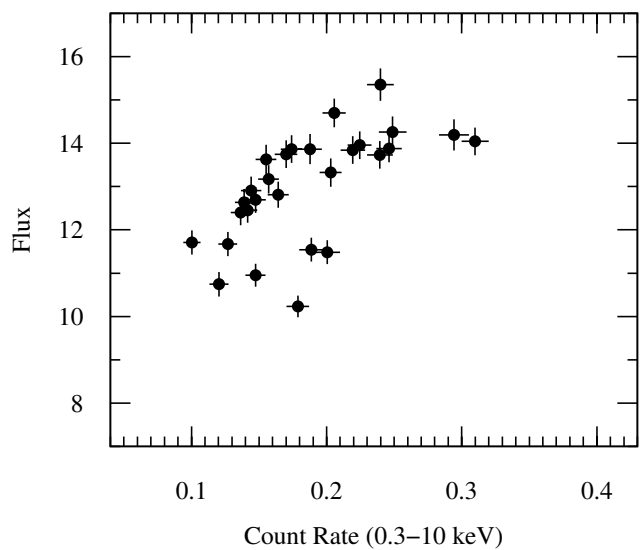}
\caption{The X-ray count rates versus the $w2$ fluxes in units of $10^{-16}$\,erg\,s$^{-1}$\,cm$^{-2}$\,\AA$^{-1}$ in 2010 October--November. 
The intervals between adjacent observations in the $w2$ and X-rays are 1--3\,days. 
The results of the Spearman rank correlation test can be found in Table~\ref{short_variability}, together with the other two UV bands. \label{xrt_uvot_short}}
\end{figure}

\begin{figure}
\epsscale{1.20}
\plotone{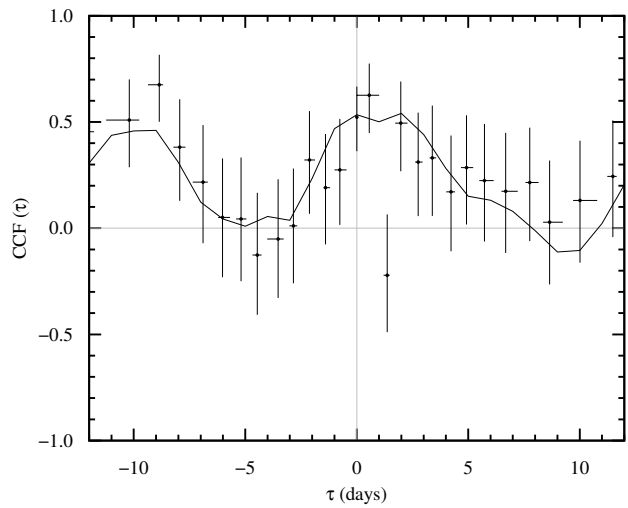}
\caption{Cross-correlation functions between the X-ray and $w2$ band during 2010 October--November: ICCF (black solid line) and ZDCF (dots with error bars). Positive time lags correspond to the X-rays leading. \label{ccf}}
\end{figure}

The light curve measured with {\it Suzaku} reveals strong variations up to a factor of two on short timescales.
To investigate possible spectral evolution with changes in the flux, 
we extract the FI spectra at the epochs of the highest and the lowest count rates, respectively, 
in the {\it Suzaku} light curve  in Figure~\ref{suzakulc} (labeled by bars).
Each spectrum has a minimum of 50 counts per bin. 
The  high- and low-flux spectra show very similar profiles,  as indicated in  Figure~\ref{high_low_spec}. 
They are then fitted with the above power law plus black body model, respectively.
Although the high-state spectrum has a slightly steeper best-fit index value than the low-state, the fitted parameters of the two spectra are consistent within the mutual uncertainties, as listed in Table~\ref{best_fit_high_low}. 
We also calculate the flux ratios of the black body to the power-law component within 0.3--10\,keV for the two spectra, and find them to be the same. 
Then we follow a different approach. 
We extract the high-state and low-state spectrum by separating them above and below the count rate threshold of 2.4 counts\,s$^{-1}$, and fit them with the above model. 
Both spectra have photon indices of $\Gamma=1.87$ and black body temperature of $kT=0.15$\,keV. 
The flux ratios of the black body to power-law component for high and low state are still the same. 
These results are consistent with the energy-independent variability as seen in Figure~\ref{suzakulc}.
We thus conclude that, on timescales as short as tens of thousand seconds and with the 
variations within a factor of two or so, there is little or no noticeable change in the spectral shape.

To investigate possible spectral variability on timescales of days, the 
{\it Swift}/XRT spectra in the  0.3--10\,keV band are extracted from each of the observations
taken during 2010 October--November,  as described in Section~\ref{swift}.
Although there is a soft excess in its X-ray spectrum, the spectral S/N are generally low, and the purpose here is to find the variation of the spectral shape possibly caused by, e.g., changes of the ratio of the soft and the hard components. 
So we fit these spectra using a single power-law model with the absorption column density fixed at the Galactic value. 
The best-fit photon indices are plotted in Figure~\ref{index}. 
No spectral variability is detected ($P=0.2$ using the $\chi^2$ test). 

\subsection{Broad Band SED}
\label{sed_section}

Given its hybrid nature, sharing properties of both NLS1s and blazars, it would be revealing to study the 
broad band SED for 1H~0323+342. 
\citet{2009ApJ...707L.142A} reported an SED 
using the averaged X-ray and UV/optical data from the {\it Swift} observations between 2006 July and 2008 November. 
It showed a low synchrotron-peak frequency at about $10^{13}$~Hz and a strong disk/corona component. 
Given the relatively large variability of the source, it would be ideal to construct an SED based on 
simultaneous observations.
Here we  make use of the X-ray data from the {\it Suzaku} observation taken on 2009 July 26--27 
and the UV/optical data taken from the {\it Swift} observation on 2009 July 27, which can be considered as 
quasi-simultaneous. 
However, the data in the other wavebands were not simultaneously taken. 
The GHz radio data are taken from the NASA/IPAC Extragalactic Database\footnote{http://ned.ipac.caltech.edu/} and listed in Table~\ref{radio_data}.
Data in the infrared, hard X-ray and $\gamma$-ray bands are retrieved from the AllWISE Source Catalog \citep{2012yCat.2311....0C}, the {\it Swift}/BAT 70-Month Catalog \citep{2013ApJS..207...19B} and the {\it Fermi}/LAT 2-year Point Source Catalog \citep{2012ApJS..199...31N}, respectively. 
The broad band SED is shown in Figure~\ref{sed}.

The broad-band SED is fitted with a simple one-zone leptonic jet model, which consists of synchrotron, synchrotron self-Compton (SSC) and external-Compton (EC) processes \citep[e.g.][]{2009MNRAS.397..985G, 2014ApJ...788..104Z, 2014arXiv1406.1934S}, 
plus emission from an accretion disk/corona. 
The $\chi^2$ minimization technique is applied to perform the fitting \citep[see][for details]{2014ApJ...788..104Z}. 
The energy distribution of the injected relativistic electrons is assumed to be a broken power law in the range of $[\gamma_{\rm min},\gamma_{\rm max}]$ with a break energy $\gamma_{\rm b}$. 
The $\gamma_{\rm max}$ is usually poorly constrained and does not affect the results significantly, so it is fixed at a large value. 
The break energy $\gamma_{\rm b}$ and the energy density parameter $N_0$ can be replaced by the peak frequency of the synchrotron component $\nu_{\rm syn}$ and the corresponding peak luminosity $\nu_{\rm syn}l_{\nu}$ \citep[see Equation (2) and (5) in][]{2009ApJ...701..423Z} during the fitting, so we set $\nu_{\rm syn}$ and $\nu_{\rm syn}l_{\nu}$ as free parameters instead of $\gamma_{\rm b}$ and $N_0$. 
The two indices $p_1$, $p_2$ of the broken power law are obtained from the spectral slopes below and above the second bump of the observed SED as described in \citet{2012ApJ...752..157Z, 2014ApJ...788..104Z}. 
Here the jet radiation region is assumed to be a sphere with a size determined by the variabillity of jet emission as adopted by \citet{2014arXiv1406.1934S} specifically for 1H~0323+342. 

One of the key factors to determine the relative importance of different sources supplying seed photons for inverse Compton is the distance of the dissipation region in the jet from the black hole, which is still controversial in AGNs \cite[e.g.][]{2009MNRAS.397..985G}. 
Some of the studies assume that the process takes place within the broad line region (BLR) and the seed photons are dominated by radiation from the BLR (hereafter IC/BLR) because of its much higher energy density than that of the torus \citep[e.g.][]{2009ApJ...699.2002B, 2013ApJ...779..100I, 2014ApJ...788..104Z}, whereas in other cases the process is assumed to takes place outside the BLR and the seed photons come predominantly from a dusty torus as infrared emission (IC/IR) since the photon field from the BLR will decrease very quickly in this case \citep[e.g.][]{2013MNRAS.435L..24T, 2012MNRAS.426..317D, 2013MNRAS.436..191D}. 
The latter is usually invoked to explain the detection of very high energy gamma-rays extending to the TeV regime in some blazars, which cannot be accounted for by the IC/BLR process due to the Klein-Nishina effect \cite[e.g.][]{2011A&A...534A..86T}. 
For 1H~0323+342, no TeV $\gamma$-ray emission is detected with a 95\% upper limit of $5.2\times10^{-12}$\,erg\,s$^{-1}$\,cm$^{-2}$ at energies above 400\,GeV \citep{2004ApJ...613..710F}. 
However, this can not rule out the IC of seed photos from a putative torus. 
Here we consider both cases that the dissipation region is inside and outside the BLR, respectively. 
In the first case, the dissipation region is inside the BLR so that IC/BLR dominates and IC/IR can be ignored. 
The energy density of the BLR in the comoving frame, $U^{'}_{\rm BLR}=1.55\times10^{-2}\Gamma_{\rm jet}^{2}$\,erg\,cm$^{-3}$, is obtained from its strong broad Balmer lines ($\sim14.5\times10^{-14}$\,erg\,s$^{-1}$\,cm$^{-2}$, Zhou 2007, private communication), calculated using the method of \citet{2014arXiv1406.1934S}, where the Lorentz factor $\Gamma_{\rm jet}$ of the jet is taken as its Doppler boosting factor, i.e., $\Gamma_{\rm jet}=\delta$, by assuming that its relativistic jet is inclined close to the line of sight given its $\gamma$-ray detection. 
Alternatively, we also consider the case where the IC is dominated by seed photons from a putative dusty torus (IC/IR). 
The energy density $U^{'}_{\rm IR}=3\times10^{-4}\Gamma_{\rm jet}^{2}$\,erg\,cm$^{-3}$ is adopted as a conventional value, as in previous works \citep[e.g.][]{2007ApJ...660..117C}. 
As shown below, both cases lead to satisfactory fits to the non-thermal emission of the broad-band SED.

Given the prominent broad emission lines in the optical spectrum as in other NLS1s, a strong ionizing continuum emission, which originates from an accretion disk around the black hole, must be present in 1H~0323+342. 
We use a multi-temperature black body model of a standard thin disk \citep{2011ApJ...728...98D} to describe this component\footnote{Given the sparse data in the optical to EUV band, we simply use the multi-temperature black body model of the geometrically thin, optically thick standard accretion disk to fit the SED. It should be noted that this model may not be appropriate for 1H~0323+342 because the derived nominal value of the Eddington ratio of this object is higher than the critical value for the geometrically thin approximation to be valid. 
Thus the fitted Eddington ratio value should be treated with caution. 
Moreover, the large uncertainty in the black hole mass based on the single-epoch spectrum estimation ($\sim0.3$\,dex) introduces a further uncertainty in the derived Eddington ratio.  
See discussion in Section\,5.1. 
}. 
A similar component was also adopted in modeling the SED of 1H~0323+342 in the previous studies \citep{2009ApJ...707L.142A, 2014arXiv1405.0715P, 2014arXiv1406.1934S}.
A black hole mass of $1.8\times10^7M_{\odot}$ determined from the broad H$\beta$ luminosity by \citet{2007ApJ...658L..13Z} is employed. 
The inner and outer radii of the disk are fixed at 3$R_{\rm sch}$ and 700$R_{\rm sch}$ respectively ($R_{\rm sch}$ is the Schwarzschild radius) and the inclination is taken to be $\cos i=1$. 
The accretion rate is varied to match the UV fluxes. 
It should be noted that the optical fluxes can only be considered as upper limits due to possible contamination from the host galaxy (marked as arrows in Figure~\ref{sed}), while the UV fluxes are not (see Section~3.2). 
For the X-ray (Suzaku) data below 10\,keV, the best-fit physical model of a power law plus Comptonization by warm plasma for the soft X-ray excess  is adopted (see Section 4.1). 

We ignore the thermal emission in the infrared band from a putative dusty torus in consideration of a few arguments. 
First, in a few similar radio-loud NLS1s, the infrared emission in the {\it WISE} bands is highly variable on the timescales of hours, 
and is considered to be dominated by non-thermal emission from the jets \citep{2012ApJ...759L..31J}.
Second, based on previous SED modeling for several $\gamma$-ray detected radio-loud NLS1s 
including  1H~0323+342, 
the contribution to the infrared from an assumed torus is found to be  much smaller  
compared with the synchrotron radiation of the jets and is negligible \citep{2009ApJ...707L.142A}. 
Last, adding a contribution in the infrared from a putative torus, which is uncertain on its own anyway, 
will have essentially little or no effect on our conclusions. 

During the fitting procedure, the free parameters $B$, $\delta$, $\nu_{\rm syn}$, $\nu_{\rm syn}l_{\nu}$ and $\gamma_{\rm min}$ are constrained with the $\chi^2$ minimization technique, while all other parameters are fixed at the given values as discussed above. 
The best-fit model parameters are listed in Table~\ref{sed_params} for both the two cases, together with their 1-$\sigma$ errors. 
We note that the UV fluxes are dominated by the thermal disk emission and the X-ray band below 10\,keV is dominated by the corona emission in both the IC/BLR and IC/IR cases, which give satisfactory fits to the SED in the infrared, and from the hard X-ray (tens of keV) to the GeV $\gamma$-ray band. 
The relative contributions of the disk/corona and the jet components in the UV/optical and X-ray bands are consistent in these two cases. 
We plot the fitting results of the IC/BLR case as a demonstration in Figure~\ref{sed}. 
As indicated in the figure, 
only in the hard X-ray band ($>10$\,keV), the jet emission starts to make a noticeable and gradually increasing contribution, and eventually dominates the observed flux at energies around 100\,keV and above. 
We also note that the infrared data are not matched well by the model, except for their overall intensities. 
This may be due to a contribution from the host galaxy in the near-infrared. 

Since the UV/optical and X-ray ($<10$\,keV) data are not simultaneous with observations in the other bands, 
the SED may be somewhat different at different epochs 
given the X-ray variability seen in this source \citep{2009AdSpR..43..889F}. 
Here we examine to what extent the above modeling will differ, due to the source variability, 
especially concerning the UV/optical and X-ray bands presented in this work. 
We vary the hard X-ray flux in the energy band of {\it Swift}/BAT (14--195\,keV)
by a factor of two in both directions, and re-do the fitting. 
These fitting results are over-plotted in Figure~\ref{sed} with grey lines. 
We find that such changes have almost no effect in the UV/optical regime, 
and the UV fluxes are always dominated by thermal emission from the accretion disk. 
In the X-ray band below 10\,keV the contribution from the jet emission is still negligible
compared with the dominant corona component. 
It should be noted that the dominance of the thermal disk and corona emission in the UV and 
X-rays below 10\,keV, respectively,  is essentially not affected by 
varying the relative contributions of the coronal X-ray luminosity, the black hole mass, 
and the jet parameters within the allowed space that match the  model SED with the data observed. 

Using our SED model, we can quantify the relative contributions from each of the components
to the observed luminosities in the {\it Swift} and {\it Suzaku} bands. 
The UV luminosity is integrated from 1300\AA~to 3300\AA, 
covering the wavebands of the three UVOT ultraviolet filters \citep{2008MNRAS.383..627P}, 
yielding $L_{\rm UV}\approx1.9\times10^{44}$\,erg\,s$^{-1}$. 
The jet component contributes only 1.3\%, $L_{\rm UV,jet}\approx2.4\times10^{42}$\,erg\,s$^{-1}$. 
The total 0.3--10\,keV luminosity is $L_{\rm X}\approx2.5\times10^{44}$\,erg\,s$^{-1}$,
 where the jet emission from inverse-Compton contributes only 
 $L_{\rm X,jet}\approx1.2\times10^{43}$\,erg\,s$^{-1}$,
i.e.\  $\sim 5\%$. 
 The fraction may be even smaller if the soft X-ray excess component is taken into account. 
We thus suggest that, in the context of the simple one-zone leptonic model, 
the jet contribution to the X-ray band ($<$10\,keV) is most likely negligible compared with the 
disk/corona emission at the epochs of the {\it Swift} and {\it Suzaku} observations considered here.

\section{SUMMARY AND DISCUSSION}
\label{discussion}

In this work, we analyze the X-ray and UV/optical data from the long-term  
monitoring with {\it Swift} of the NLS1-blazar hybrid object 1H~0323+342. 
We also analyze the {\it Suzaku} observation of this source taken on 2009 July 26--27. 
The main results are summarized as below, and their implications are discussed 
in the following subsections. 
(i) The object is variable in both the UV and X-ray bands on timescales from days to years,
and there is a statistically significant correlation between the UV flux and X-ray count rates on these timescales.
On timescales as short as a few tens of thousands seconds, X-ray variability 
by a factor of two is also found, though no significant spectral variation is detected. 
(ii) Using observations of intensive (roughly daily) monitoring with {\it Swift} lasting for $\sim$35 days, 
a cross-correlation analysis suggests a possible time lag around $\tau\sim0$~day between the UV and the X-ray emission with X-rays tentatively leading. However, the present data do not allow any firm conclusion, and
 future monitoring observations are needed to confirm this result. 
(iii) The high S/N X-ray spectrum of 1H~0323+342 taken with {\it Suzaku}/XIS is typical of Seyfert 1s, 
 a power law with a photon index of  $\Gamma\sim1.9$ and a soft excess, 
 the latter contributing to  about $11\%$ of the flux within 0.3--10\,keV. 
(iv) The broad-band SED is constructed with the (quasi-) simultaneous  UV/optical and X-ray data, 
as well as the new detection of this object in the 100--300\,MeV band with {\it Fermi}/LAT.
The latter is important for pinning down the exact position of the high-energy tail of the 
$\gamma$-ray bump (thus the energy budget of the jet emission), improving upon 
the loosely constrained limit adopted in previous SEDs. 
(v) The SED can be well modeled in the context of a simple one-zone leptonic jet  model
plus the accretion disk/corona components. 
The latter are suggested to dominate the emission in the UV and  X-ray (up to 10\,keV) bands
as observed with  {\it Swift} and {\it Suzaku} presented here.

\subsection{A NLS1-Blazar Hybrid and the UV/X-ray Emission}

The $\gamma$-ray NLS1s discovered so far are all suggested to be low synchrotron-peaked blazars 
\citep{2009ApJ...707L.142A, 2012MNRAS.426..317D}. 
Our SED fitting reveals that the synchrotron component of 1H~0323+342 peaks at $\sim10^{13}$Hz, 
consistent with this assertion \citep{2009ApJ...707L.142A}.
Considering the strong broad H$\beta$ line with an equivalent width of 
58\AA\, \citep[][]{2007ApJ...658L..13Z}, 
1H~0323+342 is similar to flat-spectrum radio quasars (FSRQs),
whose synchrotron peaks are also at low frequencies in the infrared 
 \citep{2010ApJ...716...30A, 2012A&A...541A.160G}. 
The SED fitting gives a bolometric luminosity of $L_{\rm bol}=1.9\times10^{45}$\,erg\,s$^{-1}$ (Table~\ref{sed_params}, here the bolometric luminosity is the sum of the disk and corona component). 
Adopting the  mass of the black hole $1.8\times10^7M_{\odot}$  estimated from the H$\beta$ line
\citep[][]{2007ApJ...658L..13Z}, 
the Eddington ratio is then 0.8 based on the SED modeling\footnote{We note that such a value should be treated with caution since, apart from the large uncertainty in the estimated black hole mass ($\sim0.3$\,dex for estimation based on single-epoch spectra), the actual accretion flow may differ from the standard thin disk model \citep[see, e.g.,][]{1988ApJ...332..646A}, which is used to fit the UV data here. }. 
It should be noted that Zhou et al. (2007) gave $L_{\rm bol}=1.2\times10^{45}$\,erg\,s$^{-1}$,
a factor of 1.6 lower, 
by using the bolometric correction factor for quasars of \citet{1994ApJS...95....1E}. 
By using the UV spectral slope $\alpha_{\rm UV}=1.13$ fitted from the three UV fluxes and applying the relation of $\alpha_{\rm UV}$ and the Eddington ratio for NLS1s in \citet{2010ApJS..187...64G}, we find  a  value of 0.51; however, there is a large scatter in this relation.  
 Such a high Eddington ratio is characteristic of 
typical NLS1s, and is most likely the underlying driver for the strong Fe{\scshape~ii} emission
and other properties found in 1H~0323+342, as in other NLS1s.
Increasing the black hole mass will result in a decreased Eddington ratio, and vice versa.
However, if  the observed extreme properties of NLS1s are indeed driven by
their high Eddington ratios as commonly thought, the black hole mass of 1H~0323+342 cannot be much higher 
(e.g., a mass ten times higher will lead to an Eddington ratio $<0.1$). 

The dominance of the thermal disk/corona emission in the UV and  0.3--10\,keV X-ray bands
is also consistent with previous results \citep{2009ApJ...707L.142A}. 
The jet emission contributes only a few percent to the total luminosity in these two bands (Section~\ref{sed_section}), 
though its contribution may increase toward higher energies in the hard X-ray band \citep{2009AdSpR..43..889F}. 
It should be noted that, in some other $\gamma$-ray detected NLS1s, 
the X-rays below 10\,keV  could be dominated by the jet emission, at least at some occasions
\citep[e.g., PMN J0948+0022 and PKS 2004-447,][]{2009ApJ...707L.142A}. 
This may be caused by the emergence of an inverse-Compton component in the 
X-ray band during the jet flares, which seems not to be the case for 1H~0323+342 
at the epochs of the {\it Suzaku} and {\it Swift} observations analyzed here. 

\subsection{On the Long-Term X-ray/UV Variations}

Thanks to the extensively long  monitoring observations carried out by {\it Swift},
statistically correlated variability between the UV and X-ray bands are found in 1H~0323+342,
which are  significant albeit with large scatter. 
The long-term flux variations in the optical/UV in AGNs have been studied extensively over the past decades,
for individual objects \citep[e.g.][]{1996ApJ...470..364E, 2002ApJ...580L.117M, 2013MNRAS.429...75A} and for large samples \citep[e.g.][]{1985ApJ...296..423C, 2002ApJ...564..624T, 2004ApJ...601..692V, 2013AJ....145...90A}. 
It has also been found that, in general, 
NLS1s vary with systematically smaller amplitudes compared to
AGNs with larger broad line widths (Ai et al. 2013).
The long-term UV flux variations in 1H~0323+342 (Table\,\ref{short_variability}) have
 amplitudes that agree well with the median values for the NLS1 sample in Ai et al. (2013, their Figure\,7), and
thus likely have the same origin. 
If the UV bands are dominated by the disk component during the {\it Swift} monitoring, as suggested by the SED above, the long-term UV variability can be well explained as variations 
of the accretion rate, as commonly believed to operate in AGNs \citep[e.g.][]{2008MNRAS.387L..41L}. 

The timescale of such variations, in the framework of the standard optically thick disk model for an accreting black hole, is characterized by the viscous timescale $\tau_{\rm visc}$ \citep{2006ASPC..360..265C}, 
%
\begin{equation}
	\tau_{\rm visc}=\tau_{\rm th}(R/H)^2=(\Omega \alpha)^{-1} (R/H)^2,
\end{equation}
where $\tau_{\rm th}$ is the thermal timescale, 
$\Omega$ the Kepler angular velocity, 
$\alpha$ the viscosity parameter 
and $R/H$ is the ratio of the disk radius to the scale height. 
By assuming $\alpha=0.1$, $R/H=10$, and adopting 
$M_{\rm BH}=1.8\times10^7M_{\odot}$, 
 we estimate a viscous timescale at $R=10R_{\rm sch}$ of 
 $\tau_{\rm visc}=1\times10^7{\rm~s}\approx120$~days.
The  long-term UV/optical variations as observed with {\it Swift} (Figure~\ref{swiftlc})
may be caused by such fluctuations in the mass accretion rate.
The X-ray emission from the hot corona is ultimately powered by accretion, though  the
exact physical mechanisms to transfer the accretion energy from disk to corona
 is not understood yet.
In any case, the processes must be efficient and the timescales are likely much shorter than
the viscous timescale in the disk, such as via the process of magnetic reconnection \citep{2003ApJ...587..571L}. 
In such a scenario, the correlated X-ray and UV/optical variability are expected naturally
on the timescales of several months to years. 

\subsection{On the Short-Term Correlated X-ray/UV Variations}

Now we consider  shorter timescales of about several days, on which 
1H~0323+342 also varied in the X-ray and UV bands (Figure\,\ref{w2_x} and Table\,\ref{short_variability}).  
For the  X-ray emitting corona (geometrically thick, $R/H\sim1$), the viscous timescale $\tau_{\rm visc}^{\rm c}$ is comparable to the thermal timescale $\tau_{\rm th}^{\rm c}$, 
which is much shorter than the viscous timescale of the disk. 
For the above black hole mass, 
 $\tau_{\rm visc}^{\rm c}\approx \tau_{\rm th}^{\rm c}\approx10^5$s.
 Thus the X-ray variability can be interpreted to arise from variations in the corona mass density
 and temperature. 
 The former may be related to the fluctuations in the accretion flow of the corona gas, 
 and the latter may be related to the fluctuations in the energy dissipation (heating) 
 process of the corona. 
It is thus expected for the X-rays to vary on timescales of days or longer.

The UV variability on these timescales cannot be explained by fluctuations 
of the mass accretion rate, since these timescales are much shorter than the 
viscous timescale of the disk.
Such variability is commonly seen in AGNs  from the monitoring observations
for individual objects   \citep[e.g.][]{1996ApJ...470..364E, 2002ApJ...580L.117M, 2013MNRAS.429...75A}
as well as for large samples \citep[e.g.][]{2001ApJ...555..775C}. 
On average, on timescales of days, the UV/optical variability amplitudes are small 
(typically several percent) 
 for Seyfert galaxies; they are even smaller for NLS1s based on their ensemble 
structure functions  (Ai et al. 2013, their Figure\,10), which show a rise as 
a power-law function with a slope of 0.3--0.4. 
The UV variability amplitudes found in 1H~0323+342 
during the 35-day intensive monitoring (Figure\,3 and Table\,\ref{short_variability}) are roughly comparable 
to those of the Seyfert 1 population. 
The physical mechanism governing the short-term UV/optical variations  is not well understood yet, however. 

One of the most interesting results of this work is the evidence for statistically correlated UV--X-ray variability on timescales of days. 
This is observed for the first time among $\gamma$-ray detected NLS1s. 
Similar correlated UV/optical--X-ray variations on similar timescales have been previously reported in 
a limited number of AGNs, mostly radio-quiet, 
with intensive simultaneous monitoring observations, 
 e.g., NGC 4051 \citep{2010MNRAS.403..605B,2013MNRAS.429...75A}, MR2251-178 \citep{2008MNRAS.389.1479A}, Mrk79 \citep{2009MNRAS.394..427B}, PG 1211+143 \citep{2009MNRAS.399..750B}. 
The most promising explanation invokes 
reprocessing of the primary X-ray emission by an irradiated `cold' accretion disk, 
whose local temperature may, in turn, be altered by the reprocessed energy dissipated into the disk. 
Changes in the irradiating X-ray emission would cause subsequent changes of the
local emissivity in the disk, leading to the corresponding variations in the UV/optical emission \citep{2003A&A...400..437C}. 
The reprocessing model has been used to explain the  UV/optical variations
on the timescales of days  and their correlated X-ray variations  observed in several  AGNs 
\citep[e.g.][]{1992ApJ...393..113C, 1998ApJ...505..594N, 2000ApJ...544..734N, 2001ApJ...561..162S, 2002ApJ...580L.117M, 2008MNRAS.389.1479A, 2009MNRAS.394..427B, 2010MNRAS.403..605B, 2013MNRAS.429...75A}. 

Based on the argument of energetics, the changes in the UV/optical luminosity 
 should always be smaller than (or at most comparable to) the changes in the X-ray luminosity, 
which is the driving component. 
In the case of 1H~0323+342, during the 35-day monitoring in 2010 October--November, the amplitude of the luminosity variations in X-ray band is $\Delta L_{\rm X}=6.1\times10^{43}$\,erg\,s$^{-1}$, compared to $\Delta L_{\rm UV}<1.9\times10^{43}$\,erg\,s$^{-1}$ in the UV band. 
The fractional variability amplitude in the X-ray band, $F_{\rm var,X}=29\%$, is also larger than  that in the UV,  $F_{\rm var,UV}<10\%$. 

When a variation in the driving X-ray emission occurs, a time lag is expected for the  UV/optical emission to respond, which is roughly the readjustment timescale  (dynamical timescale) of the surface layer of the disk, heated by the irradiating X-ray emission. 
Such a timescale is generally short. 
As discussed in \citet{2003A&A...400..437C},  assuming that the Thomson thickness 
is unity and the readjustment takes place at the sound speed, this timescale for an AGN with $10^8$ $M_{\odot}$ black hole mass is: 
\begin{equation}
	t_{\rm dyn}\sim10^5 T^{-1/2}_{6}n_{12}^{-1}\,{\rm s},
\end{equation}
where $T_6$ is the disk temperature in units of $10^6$\,K and $n_{12}$ the electron number density 
in units of $10^{12}$\,cm$^{-3}$. 
For 1H~0323+342 with a smaller black hole mass, this timescale could be $<10^5$~s.
The time lag of around zero days with a possible leading by the X-rays, as indicated by the cross-correlation analysis in Section~\ref{correlated_uv_x},  is consistent with this estimate\footnote{The light travel time from the corona to the disk  is generally negligible,  of the order $\sim10^3$~s assuming a scale-height of the corona as  a few Schwarzschild radii above the disk.}. 
More intensive monitoring with even higher cadence of data sampling than the current observations
is needed to confirm this result, however.

Alternatively, in light of the fact that the observed time lag is consistent with zero, 
the correlated X-ray and UV emission may also be caused 
by the following effect.
As one can see from Figure 4, the variation in the full 0.2--10\,keV band is dominated by
that in the soft X-rays in 0.2--2\,keV, where the soft X-ray excess dominates.
It has been shown in Section\,4.1 that the soft X-ray excess can well be modeled as
the high-energy tail of Comptonized emission 
of seed photons from the underlying accretion disk by
warm plasma of 0.28\,keV with an optical depth of
$\tau\sim$13.4 (i.e. the mean number of scatterings $=\tau^2\sim180$).
If this is indeed the case,  the observed UV emission may also be  contributed by Comptonized photons, 
which arise either from seed photons of low energies and/or from experiencing only a small number of scatterings. 
Fluctuations in the warm plasma may give rise to synchronized variations in both the
(soft) X-ray and UV bands naturally with no time lag.

In reality, the interplay between the X-ray and UV/optical variations must be complex
rather than a simple one-to-one follow-up, considering the feedback of the UV/optical radiation to the X-rays 
as seed photons for Compton scattering, as well as other possible factors. 
For example, the observed variations of the X-rays may be caused by
small changes in the ionization state  or the covering factor of a (dust-free) ionized absorber, 
whereas the UV/optical emission would be left unaffected.
This may explain the mis-match between the X-rays and UV seen 
in the latter half of the light curves in Figure~\ref{w2_x}. 

Finally, based on our SED modeling, 
 the jet contributes only $\lesssim5\%$ to the UV and 0.3--10\,keV X-ray band. 
Therefore, though the jet contributions cannot be ruled out completely in the hard X-ray band, 
it is unlikely that the observed variations in the UV and X-ray band are significantly contributed by the non-thermal radiation from the jet, i.e., 
variations of the synchrotron and SSC components respectively, 
which are correlated by themselves
\footnote{
Shortly before the submission of
our paper, a paper by \citet{2014arXiv1405.0715P} appeared on the astro-ph preprint server. 
Compared to their work, we reached similar conclusions regarding the Swift light curve, but focus on different temporal epochs. 
We note that they reported the {\it Fermi} $\gamma$-ray light curve
and concluded that the $\gamma$-ray flux of this source was nearly stable throughout years until the end of 2012, 
when anomalous variations in all the bands including the $\gamma$-ray band were detected. 
The data we present in this work do not cover that period of time. 
Our result supports the idea that 1H~0323+342 likely spent most of the time 
(at least during 2010 October--November) in quiescence, in which 
its UV/optical and X-ray emission is dominated by the disk/corona component. 
}.

\subsection{Constraining the Black Hole Mass}
\label{excessvariance}

Given their relatively small widths of the broad lines, 
the black hole masses in NLS1s are believed to be systematically lower than those in classical Seyfert 1s and quasars \citep[e.g.][]{2004ApJ...606L..41G, 2004AJ....127.3168B, 2012AJ....143...83X}
estimated from the commonly used virial method. 
This, and the resulting high Eddington ratios \citep[e.g.][]{2010ApJS..187...64G, 2012AJ....143...83X}
can explain  naturally  most of the observed properties of NLS1s (see \citealt{2008RMxAC..32...86K} for a review). 
Major properties of NLS1s include their location at one extreme end
in the so-called Eigenvector-I parameter space \citep{2002ApJ...565...78B}, 
rapid X-ray variability \citep[e.g.][]{1999ApJS..125..297L, 2011ApJ...727...31A}, 
relatively low amplitudes of  optical/UV variability \citep[e.g.][]{2013AJ....145...90A}, 
strong gaseous outflows as traced by the [OIII]5007 emission \citep[e.g.][]{2007ApJ...667L..33K}, 
and the morphology (pseudo-bulges) of their host galaxies \citep[e.g.][]{2011MNRAS.417.2721O, 2012ApJ...754..146M}. 
However, there remains controversy as to whether their black hole masses are underestimated,
if their broad line regions are planar and seen face on 
\citep{2006MNRAS.369..182J}. 
Specifically, with arguments along the same line, 
\citet{2013MNRAS.431..210C} proposed recently that radio-loud NLS1s have supermassive 
black holes similar to those in blazars ($\sim10^8-10^9M_{\odot}$). 

An  independent constraint on the black hole masses of AGNs comes from the variability of their X-ray emission, 
which carries the information about the dynamics of the hot corona in the close vicinity of the black holes. 
The {\it Suzaku} observation  shows that 1H~0323+342 varied significantly on a timescale of 
$\sim$20~ks (Figure~\ref{suzakulc}). 
Short timescales as such are almost ubiquitous to AGNs with relatively small black hole masses \citep{1999ApJS..125..297L, 2009MNRAS.394..443M, 2011ApJ...727...31A, 2014ApJ...782...55Y}. 
This minimum variability timescale sets an upper limit on the light-crossing time of the X-ray emitting region with a typical size of  $\sim$10 Schwarzschild radii, 
which corresponds to a black hole mass  $<4\times10^{7}M_{\odot}$.

A more quantitative estimation can be achieved by using the X-ray normalized excess variance 
$\sigma^2_{\rm rms}$, defined as the variance of the light curves divided by the mean flux with the contribution from the measurement errors subtracted 
\citep[e.g.][]{1997ApJ...476...70N, 2003MNRAS.345.1271V}. 
Interesting enough, a remarkably tight linear relationship between the 
short-timescale X-ray normalized excess variances  and the black hole masses
was suggested in recent studies by 
\citet{2005MNRAS.358.1405O}, \citet{2010ApJ...710...16Z} and \citet{2012A&A...542A..83P}. 
This provides a useful tool to estimate the black hole masses independent of the commonly used virial method. 
To make it directly comparable to previous results and also to eliminate any potential contribution from the jet at high energies, we extract the 2--4\,keV light curve\footnote{
It should be noted that this relation is insensitive to the energy band of the X-ray  emission within which $\sigma^2_{\rm rms}$ is calculated. 
Extracting a light curve in the 0.2--2\,keV band gives $\sigma^2_{\rm rms,0.2-2}=17.1^{+11.2}_{-5.2}\times10^{-3}$, leading to a slightly smaller $M_{\rm BH}$. 
}
 from the {\it Suzaku} observation and divide it into four segments of 40~ks each. 
 The normalized excess variance and its uncertainty is calculated following \citet{2012A&A...542A..83P}. 
 We find an estimation of $\sigma^2_{\rm rms,2-4}=12.3^{+8.1}_{-3.9}\times10^{-3}$, which corresponds to a black hole mass of $M_{\rm BH}=8.6^{+2.9}_{-2.7}\times10^6M_{\odot}$ using the relation obtained from
the AGN sample with black hole masses estimated via the reverberation mapping method \citep{2012A&A...542A..83P}. 
This value is consistent with that estimated from the broad H$\beta$ line \citep[$\sim1.8\times10^7M_{\odot}$,][]{2007ApJ...658L..13Z}, 
whose uncertainty is 0.3--0.4 dex. 
In Figure~\ref{var},
we  compare the obtained $\sigma^2_{\rm rms}$ and the $M_{\rm BH}$
estimated from the broad line width of H$\beta$ for 1H~0323+342 with 
 the $\log \sigma^2_{\rm rms}$--$\log M_{\rm BH}$ relation. 
An error bar indicating the typical uncertainty of $M_{\rm BH}$ (0.3 dex) is also plotted. 
As can be seen, the two methods give  broadly consistent 
$M_{\rm BH}$ values  within the uncertainties. 
We conclude that, if the observed X-ray variability is indeed predominantly contributed by the emission from the corona instead of from the jet, as is suggested in this work, 
1H~0323+342  harbors most likely a small black hole with the mass 
of the order of $10^7M_{\odot}$.

\acknowledgments

We acknowledge Jin Zhang and Xiaona Sun for help with the SED modeling. 
We also thank Junjie Mao and Haiwu Pan for helpful discussion on cross-correlation analysis. 
This work was supported by NSFC Grant No. 11033007 and Grant No. 11273027, and by the Strategic Priority Research Program ``The Emergence of Cosmological Structures'' of the Chinese Academy of Sciences (Grant No. XDB09000000). 
This research has made use of data and software provided by the High Energy Astrophysics Science Archive Research Center (HEASARC), which is a service of the Astrophysics Science Division at NASA/GSFC and the High Energy Astrophysics Division of the Smithsonian Astrophysical Observatory,  
the XRT Data Analysis Software (XRTDAS) developed under the responsibility of the AOS Science Data Center (ASDC), Italy, 
and the data obtained from the {\it Suzaku} satellite, a collaborative mission between the space agencies of Japan (JAXA) and the USA (NASA). 
This research has also made use of the NASA/IPAC Extragalactic Database (NED) which is operated by the Jet Propulsion Laboratory, California Institute of Technology, under contract with the National Aeronautics and Space Administration.

\appendix

\section{Testing the Significance of the Cross-Correlation Function}
\label{append}

\begin{figure}
\epsscale{1.20}
\plotone{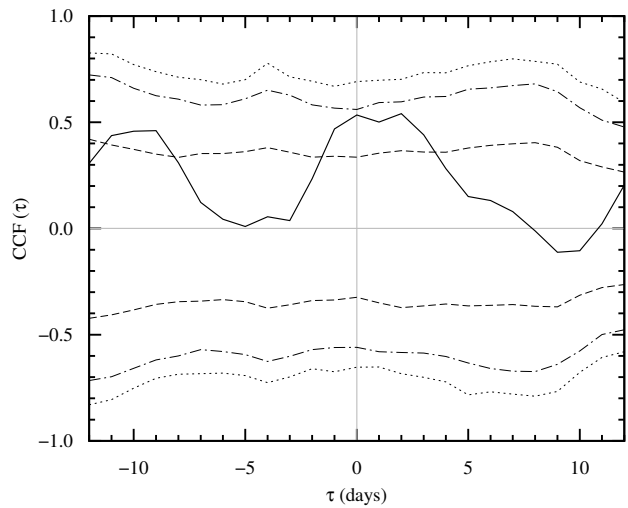}
\caption{
The ICCF between X-ray and $w2$ band light curves during 2010 October--November is plotted as black solid line. Positive time lags correspond to X-rays leading. The dashed, dot-dashed and dotted lines represent 68\%, 95\% and 99\% extremes of the distribution of the ICCFs between simulated X-ray light curves and the observed $w2$ light curve.
\label{iccf_test} 
}
\end{figure}

\begin{figure}
\epsscale{1.20}
\plotone{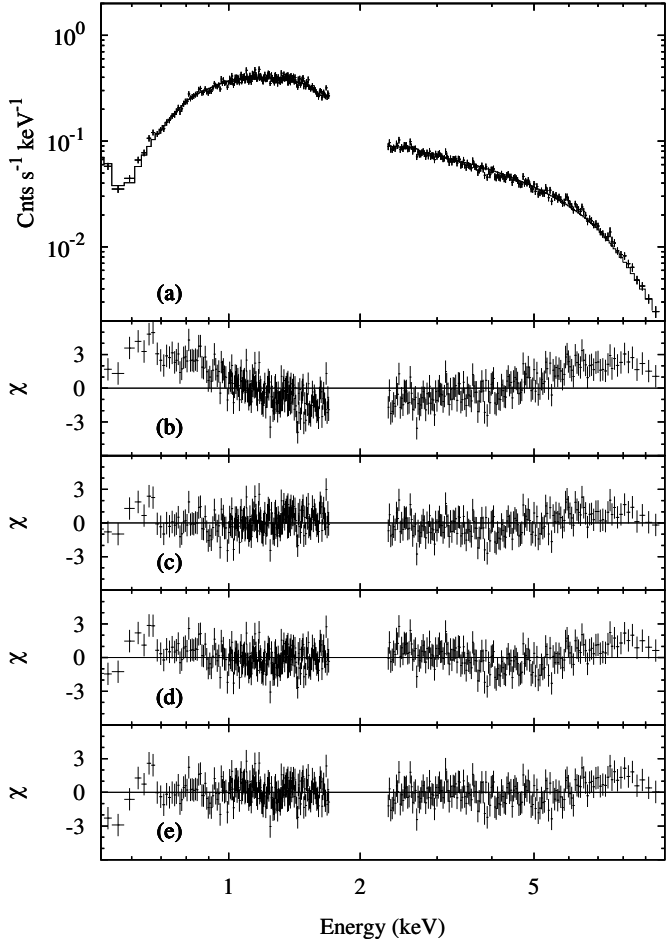}
\caption{Panel (a) shows the averaged {\it Suzaku} FI spectrum of 1H 0323+342 (crosses) and the power law plus black body model (solid line) fitted to the spectrum. 
Lower panels show the residuals of the spectral fits in terms of $\sigma$ with error bars of size one. 
Panel (b): Residuals of single power law fitted to the spectrum. 
The positive residuals are revealed below 1\,keV. 
Panel (c): Residuals of power law plus black body model fitted to the spectrum. 
Panel (d): Residuals of the ionized partial covering absorption model fitted to the spectrum. 
Panel (e): Residuals of reflection model. 
The data between 1.7--2.3\,keV are ignored on account of possible calibration uncertainties (see Section~\ref{suzaku}). 
\label{averaged_spec}}
\end{figure}

\begin{figure}
\epsscale{1.20}
\plotone{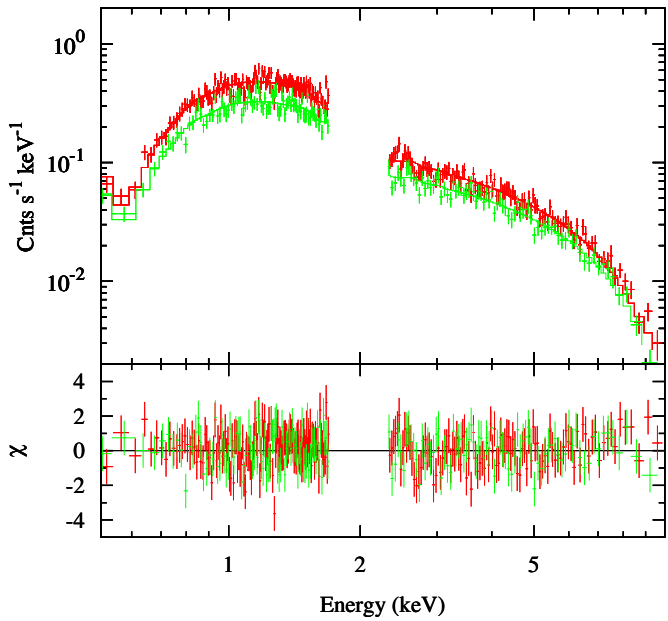}
\caption{1H~0323+342 FI spectra and their best-fit power law plus black body models in high-state (red) and low-state (green), respectively, during the {\it Suzaku} observation. 
See Figure~\ref{suzakulc} for the time intervals of high- and low-state. 
The spectral shape does not change, though the flux varied. 
See the electronic edition of the Journal for a color version of this figure.\label{high_low_spec}}
\end{figure}

\begin{figure}
\epsscale{1.20}
\plotone{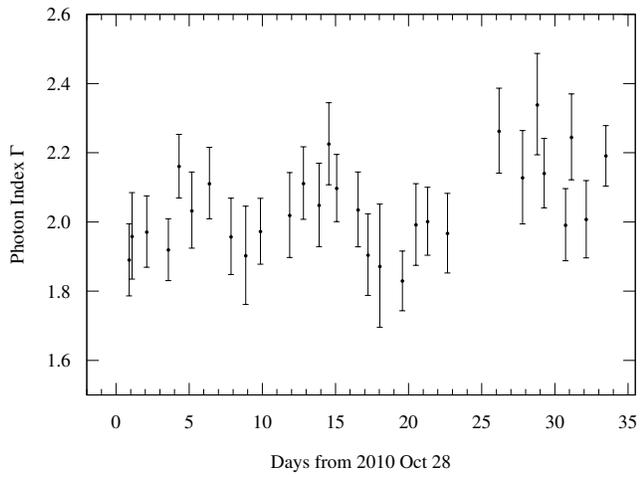}
\caption{The variations of the photon indices $\Gamma$ of the {\it Swift}/XRT observations during 2010 October--November. 
The spectra are fitted by a single power law with the absorption column density fixed at the Galactic value. 
\label{index}}
\end{figure}

\begin{figure}
\epsscale{1.20}
\plotone{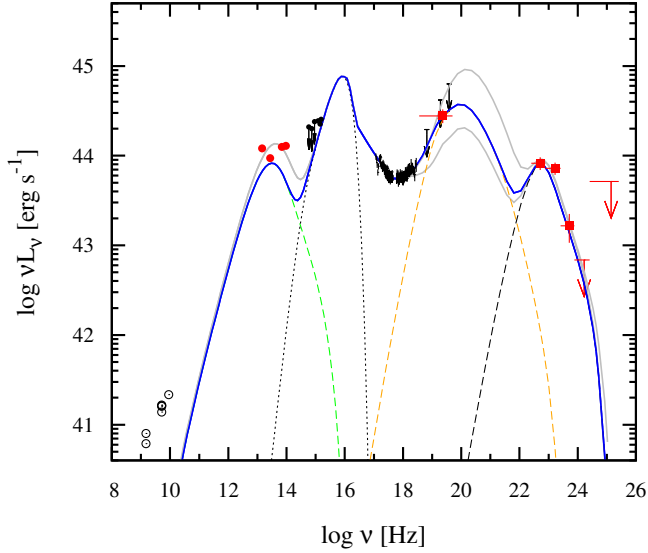}
\caption{The rest-frame broad band SED of 1H~0323+342. The red circles represent the infrared data from the AllWISE Source Catalog. 
The black arrows indicate the 90\% upper limits from the {\it Suzaku}/HXD and the red arrows indicate the 2-$\sigma$ upper limits from the {\it Fermi}/LAT. 
The UV/optical data are corrected for reddening by Galactic dust and the X-ray data are corrected for neutral interstellar absorption. 
The radio data are not involved in the SED fitting. 
The green, orange and black dashed lines represent the synchrotron, SSC and EC components, respectively. 
The dotted lines refer to the multi-temperature black body of a standard disk. 
The blue solid curve is the sum of all contributions from the best-fit model. 
The black hole mass employed here is $1.8\times10^7M_{\odot}$. 
The grey solid lines represent the SED model fitting results after the hard X-rays were varied by a factor of two. 
See the electronic edition of the Journal for a color version of this figure.\label{sed}}
\end{figure}

\begin{figure}
\epsscale{1.20}
\plotone{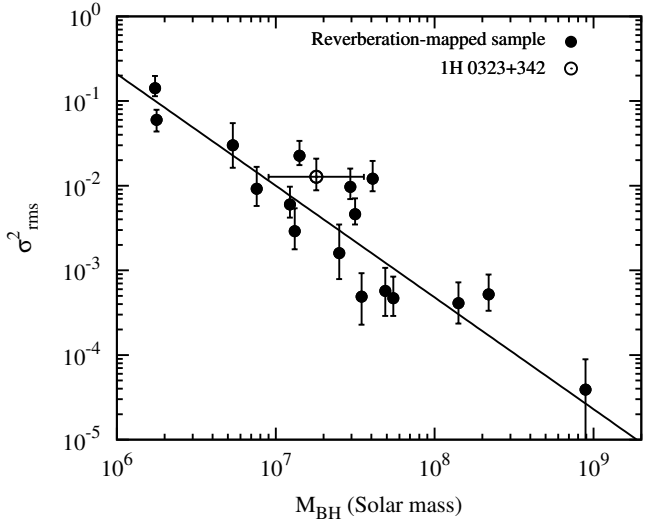}
\caption{The $\log M_{\rm BH}$-$\log \sigma^2_{\rm rms}$ relation calibrated by the reverberation-mapped sample in \citet{2012A&A...542A..83P} (solid line). 
The filled circles indicate sources in that sample and the open circle is 1H~0323+342. 
The errors on the normalized excess variances are at 1-$\sigma$ confidence level. 
A typical uncertainty (0.3 dex) of the black hole mass measured by the virial method is also plotted for 1H~0323+342. 
The sources with only upper limits in \citet{2012A&A...542A..83P} are not shown here. 
The black hole masses of sources in the reverberation-mapped sample were determined using the reverberation mapping method. \label{var}}
\end{figure}

The Monte Carlo technique is employed to generate the simulated X-ray light curves. 
First, we interpolate the observed X-ray light curve into bins of 0.1 day and calculate the power spectrum of the interpolated light curve. 
We generate the red noise light curves according to the method of \citet{1995A&A...300..707T} with the obtained power spectrum. 
These generated light curves with bins of 0.1 day are then sampled in a fashion that mimics the real observations. 
The Poisson noise is added to each point, considering the contribution from the observational uncertainty. 
In this way, the uncertainties which arise from the stochastic property of the red noise and the measurement are both included in the simulation. 
By repeating this process for 2000 times, we have 2000 simulated light curves which are uncorrelated with the observed one. 
Then we calculate the ICCF between these simulated light curves and the $w2$ light curve exactly as done for the observed X-rays. 
The values of each ICCF are recorded so that there is a distribution of ICCF values at each $\tau$. 
The 68\%, 95\% and 99\% extremes of the distributions are then plotted in Figure~\ref{iccf_test} as dashed, dash-dotted and dotted lines, respectively.




\bibliographystyle{apj}
\bibliography{references}

\clearpage

\begin{table}
\begin{center}
\caption{Results of the UVOT calibration procedure as described in Section~\ref{swift}  \label{swift_uvot}}
\begin{tabular}{lccc}
\tableline\tableline
Band & $\lambda_{\rm eff}$ & CF & A$_{\Lambda}$ \\
         & (\AA)                & (erg\,s$^{-1}$\,cm$^{-2}$\AA$^{-1}$) & (mag) \\
\tableline
$v$   & 5427   &  $2.60\times10^{-16}$  &  0.65  \\
$b$   & 4355   &  $1.46\times10^{-16}$  &  0.86  \\
$u$   & 3473   &  $1.64\times10^{-16}$  &  1.04  \\
$w1$ & 2611   & $4.31\times10^{-16}$  &  1.48  \\
$m2$ & 2255   & $8.30\times10^{-16}$  &  1.79  \\
$w2$ & 2082   & $5.86\times10^{-16}$  &  1.72  \\
\tableline
\end{tabular}
\end{center}
\end{table}

\clearpage

\begin{table}
\begin{center}
\caption{UV and X-ray Variability Amplitudes and the Results of Spearman's Rank Correlation Test \label{short_variability}}
\begin{tabular}{cccccc}
\tableline\tableline
Band &  & $F_{\rm var}$ & $\rho$ & $P$ \\
         &  & (\%) \\
(1)     &  & (2) & (3) & (4) \\
\tableline
\multirow{2}{*}{$w1$} &  & 14.9$\pm$0.4 & 0.66 & $2.8\times10^{-6}$ \\
           &                &  4.8$\pm$0.9 & 0.77 & 0.7\% \\
\multirow{2}{*}{$m2$} &  &  15.7$\pm$0.5 & 0.75 & $1.3\times10^{-7}$ \\
           &                &  7.0$\pm$1.1 & 0.88 & 0.1\% \\
\multirow{2}{*}{$w2$} &  &  14.3$\pm$0.3 & 0.75 & $<10^{-8}$ \\
           &                &  9.3$\pm$0.4 & 0.73 & $1\times10^{-5}$ \\
\multirow{2}{*}{X-ray} &  &  33.9$\pm$0.6 & $\cdots$ & $\cdots$ \\
          &   &  28.8$\pm$0.8 & $\cdots$ & $\cdots$ \\
\tableline
\end{tabular}
\tablecomments{The $\sigma_{\rm m}$, $F_{\rm var}$, $\rho$ and $p$ in the first line of each band are values for the whole dataset of the {\it Swift} observations used in this paper, and in the second line are values for the observations during 2010 October--November. 
Column 2 is the fractional variability amplitude calculated using equation~(\ref{frac_var}); 
Column 3 and 4 are the Spearman's rank correlation coefficient and the corresponding $p$-value. 
}
\end{center}
\end{table}

\clearpage

\begin{table}
\begin{center}
\caption{Parameters of Four Models Applied to the Time-Averaged Spectrum of the {\it Suzaku} Data \label{best_fit_averaged}}
\begin{tabular}{cccccc}
\tableline\tableline
\multicolumn{5}{c}{Power law+black body} \\
\tableline
$\Gamma$ & $kT$ [keV] & & & $\chi^2$/d.o.f. \\
$1.87\pm0.02$ & $0.15\pm0.01$ & & & 349/357 \\
\tableline
\multicolumn{5}{c}{Double Power law} \\
\tableline
$\Gamma_{1}$ & Norm\tablenotemark{a} & $\Gamma_{2}$ & Norm & $\chi^2$/d.o.f. \\
$3.13_{-0.17}^{+0.19}$ & $2.49\pm0.36$ & $1.58_{-0.09}^{+0.08}$ & $1.91_{-0.35}^{+0.33}$ & 319/357 \\
\tableline
\multicolumn{5}{c}{Power law$+$Partial Covering Absorption} \\
\tableline
$\Gamma$ & $N_{\rm H}$\tablenotemark{b} & $\log\xi_{\rm A}$\tablenotemark{c} & $f$\tablenotemark{d} & $\chi^2$/d.o.f.  \\
$2.40\pm0.04$ & $6.06_{-0.49}^{+0.46}$ & $-0.76\pm0.63$ & $0.58\pm0.03$ & 391/356 \\
\tableline
\multicolumn{5}{c}{Power law$+$Comptonization} \\
\tableline
$\Gamma$ & $kT_{\rm plasma}$ [keV] & $\tau$\tablenotemark{e} & & $\chi^2$/d.o.f.  \\
$1.78^{+0.04}_{-0.12}$ & $0.28_{-0.06}^{+0.56}$ & $13.4^{+4.7}_{-7.0}$ & & 330/359 \\
\tableline
\multicolumn{5}{c}{Power law$+$Reflection} \\
\tableline
$\Gamma$ & $A_{\rm Fe}$\tablenotemark{f}  & $\xi_{\rm R}$\tablenotemark{g}  & $a$\tablenotemark{h}  & $\chi^2$/d.o.f. \\
1.95$\pm$0.01 & $0.92_{-0.19}^{+0.63}$ & $225_{-14}^{+16}$ & $<0.13$ & 351/355 \\
\tableline
\end{tabular}
\tablenotetext{a}{Flux density of the power-law component at 1\,keV in units of 10$^{-3}$ photons\,keV$^{-1}$\,cm$^{-2}$\,s$^{-1}$. 
}
\tablenotetext{b}{Absorption column density $N_{\rm H}$ of the partial covering absorber in units of $10^{22}$\,cm$^{-2}$. }
\tablenotetext{c}{Ionization parameter of ionized partial covering model, defined in \citet{2008MNRAS.385L.108R}. 
}
\tablenotetext{d}{Covering fraction of the absorber. }
\tablenotetext{e}{Optical depth. }
\tablenotetext{f}{Abundance of iron relative to solar value. }
\tablenotetext{g}{Ionization parameter of reflection model, defined in \citet{2005MNRAS.358..211R}. }
\tablenotetext{h}{Spin parameter of the reflection model. The model cannot constrain this parameter and only gives an upper limit with 90\% confidence level. }
\tablecomments{The neutral absorption column density is fixed at the Galactic value $N^{\rm Gal}_{\rm H}=1.27\times10^{21}$\,cm$^{-2}$. 
}
\end{center}
\end{table}

\clearpage

\begin{table}
\begin{center}
\caption{Parameters of the Absorbed Power Law Plus black body Model Applied to the High and Low State {\it Suzaku} Spectrum 
\label{best_fit_high_low}}
\begin{tabular}{lcccc}
\tableline\tableline
 & $\Gamma$ & $kT$ [keV] & $\chi^2$/d.o.f. & $R$\tablenotemark{a} \\
 \tableline
High-state & $1.90^{+0.05}_{-0.06}$ & $0.15\pm0.02$ & 214/232 & 11\% \\
Low-state & $1.80\pm0.06$ & $0.15\pm0.02$ & 153/170 & 11\% \\
\tableline
\end{tabular}
\tablenotetext{a}{The 0.3--10\,keV flux ratio of black body to power-law component in the model. }
\tablecomments{The neutral absorption column density is fixed at the Galactic value $N^{\rm Gal}_{\rm H}=1.27\times10^{21}$\,cm$^{-2}$. 
}
\end{center}
\end{table}

\clearpage

\begin{table}
\begin{center}
\caption{The GHz Radio Data of the SED of 1H~0323+342 Retrieved from NED \label{radio_data}}
\begin{tabular}{c c c}
\tableline\tableline
Frequency  &  Flux  & References \\
 (GHz)  &  (mJy)  \\
\tableline
8.4    &  277      &  1 \\
5       &  357      &  2  \\
4.89  &  304      &  3 \\
4.85  &  368      &  4 \\
4.85  &  362      &  5 \\
4.83  &  358      &  6 \\
1.4    &  474      &  7 \\
1.4    &  614.3   &  8 \\
\tableline
\end{tabular}
\tablerefs{
(1) Healey et al. 2007; \nocite{2007ApJS..171...61H}
(2) Linford et al. 2012; \nocite{2012ApJ...744..177L}
(3) Laurent-Muehleisen et al. 1997; \nocite{1997A&AS..122..235L}
(4) Gregory \& Condon 1991; \nocite{1991ApJS...75.1011G}
(5) Becker, White, \& Edwards 1991; \nocite{1991ApJS...75....1B}
(6) Griffith et al. 1990; \nocite{1990ApJS...74..129G}
(7) White \& Becker 1992; \nocite{1992ApJS...79..331W}
(8) Condon et al. 1998. \nocite{1998AJ....115.1693C}
}
\end{center}
\end{table}

\clearpage

\begin{table}
\scriptsize
\begin{center}
\caption{Parameters of the Best-Fit SED Model Parameters in the IC/BLR and IC/IR Cases\label{sed_params}}
\begin{tabular}{lcc}
\tableline\tableline
Parameter & IC/BLR & IC/IR \\
(1) & (2) & (3) \\
\tableline
$\gamma_{\rm min}$\tablenotemark{a} &  130$\pm$15 & 28$\pm$6  \\
$\gamma_{\rm b}$\tablenotemark{b} &  1073$\pm$412  & 296$\pm$90 \\
$\gamma_{\rm max}$\tablenotemark{c} & 8000 & 12000 \\
$p_1$\tablenotemark{d} &  -1.8  & 1.8 \\
$p_2$\tablenotemark{e} &  5.0  & 4.6 \\
$\delta$\tablenotemark{f} & 2.7$\pm$0.6 & 13.6$\pm$0.9 \\
$B$\tablenotemark{g} & 1.9$\pm$0.7 & 0.3$\pm$0.25 \\
$M_{\rm BH}$\tablenotemark{h} & $1.8\times10^7$ & $1.8\times10^7$\\
$\nu_{\rm syn}$\tablenotemark{i} &  $(2.0\pm1.6)\times10^{13}$  & $(1.8\pm1.5)\times10^{12}$\\
$\nu_{\rm syn}l_{\nu}$\tablenotemark{j} &  $(1.9\pm1.7)\times10^{44}$  & $(1.1\pm0.7)\times10^{44}$ \\
$L_{\rm bol}$\tablenotemark{k} & $1.9\times10^{45}$ & $1.9\times10^{45}$ \\
$P_{\rm r}$\tablenotemark{l} &  $7.1\times10^{43}$ & $9.2\times10^{42}$  \\
$P_{\rm e}$\tablenotemark{m} &  $2.7\times10^{43}$ & $5.4\times10^{43}$ \\
$P_{\rm p}$\tablenotemark{n} &  $4.6\times10^{43}$ & $9.6\times10^{44}$  \\
$P_{B}$\tablenotemark{o} &  $4.3\times10^{42}$ & $7.7\times10^{43}$  \\
\tableline
\end{tabular}
\tablenotetext{a}{Minimum Lorentz factor of the injected electrons. }
\tablenotetext{b}{Break Lorentz factor of the injected electrons. }
\tablenotetext{c}{Maximum Lorentz factor of the injected electrons. }
\tablenotetext{d}{Slope of the injected electron distribution below $\gamma_{\rm b}$. }
\tablenotetext{e}{Slope of the injected electron distribution above $\gamma_{\rm b}$. }
\tablenotetext{f}{Doppler boosting factor. }
\tablenotetext{g}{Magnetic field in units of Gauss. }
\tablenotetext{h}{Black hole mass in units of solar mass $M_{\odot}$. }
\tablenotetext{i}{Peak frequency of the synchrotron component in units of Hz. }
\tablenotetext{j}{{L}uminosity at peak frequency of the synchrotron component in units of erg\,s$^{-1}$. }
\tablenotetext{k}{Bolometric luminosity $L_{\rm bol}=L_{\rm disk}+L_{\rm corona}$ in units of erg\,s$^{-1}$. }
\tablenotetext{l}{Radiative power in units of erg\,s$^{-1}$. }
\tablenotetext{m}{Power in the bulk motion of electrons of the jet in units of erg\,s$^{-1}$. }
\tablenotetext{n}{Power in the bulk motion of protons of the jet in units of erg\,s$^{-1}$. }
\tablenotetext{o}{Poynting power in units of erg\,s$^{-1}$. }
\end{center}
\end{table}

\clearpage

\clearpage

\clearpage

\clearpage

\clearpage

\clearpage

\clearpage

\clearpage

\clearpage

\clearpage

\clearpage

\clearpage

\end{document}